\documentclass[aps,pra,twocolumn,superscriptaddress]{revtex4-1}

\usepackage{amssymb}
\usepackage{amsmath}
\usepackage{graphicx}
\usepackage{color}
\usepackage{verbatim}
\usepackage[english]{babel}
\usepackage[latin1]{inputenc}
\usepackage{subfigure}
\usepackage{float}
\definecolor{darkblue}{rgb}{0.,0.24,0.51}
\definecolor{britishracinggreen}{rgb}{0.0, 0.26, 0.15}
\definecolor{darkgreen}{rgb}{0,0.60,.2}
\usepackage[normalem]{ulem}   

\usepackage{hyperref}
\hypersetup{
   pdfpagemode=None,
   pdfstartpage=1,
   pdfmenubar=true,
   pdftoolbar=true,
   colorlinks = true,
   linkcolor=blue,
   citecolor=blue,
   urlcolor=blue,
   bookmarksopen=false
 }
 
\newcommand{\beq}{\begin{equation}}
\newcommand{\eeq}{\end{equation}}

\newcommand{\C}{\mathcal{C}}

\newcommand{\sign}{\text{sign}}
\newcommand{\bk}{\textbf{k}}

\begin{document}

\title{Linking topological features of the Hofstadter model to optical diffraction figures}

\author{Francesco Di Colandrea}
\thanks{These two authors contributed equally}
\affiliation{Dipartimento di Fisica, Universit\`{a} di Napoli Federico II, Complesso Universitario di Monte Sant'Angelo, Via Cintia, 80126 Napoli, Italy}
\author{Alessio D'Errico}
\thanks{These two authors contributed equally}
\affiliation{Dipartimento di Fisica, Universit\`{a} di Napoli Federico II, Complesso Universitario di Monte Sant'Angelo, Via Cintia, 80126 Napoli, Italy}
\author {Maria Maffei}
\affiliation{Universit\'e  Grenoble  Alpes,  CNRS,  Grenoble  INP,  Institut  N\'eel,  38000  Grenoble,  France}
\author {Hannah M. Price}
\affiliation{School of Physics and Astronomy, University of Birmingham, Edgbaston, Birmingham B15 2TT, United Kingdom}
\author {Maciej Lewenstein}
\affiliation{ICFO -- Institut de Ciencies Fot\'oniques, The Barcelona Institute of Science and Technology, 08860 Castelldefels (Barcelona), Spain}
\affiliation{ICREA, Pg.~Llu\"is Companys 23, 08010 Barcelona, Spain}
\author{Lorenzo Marrucci}
\affiliation{Dipartimento di Fisica, Universit\`{a} di Napoli Federico II, Complesso Universitario di Monte Sant'Angelo, Via Cintia, 80126 Napoli, Italy}
\affiliation{CNR-ISASI, Institute of Applied Science and Intelligent Systems, Via Campi Flegrei 34, 80078 Pozzuoli (NA), Italy}
\author{Filippo Cardano}
\email{filippo.cardano2@unina.it}
\affiliation{Dipartimento di Fisica, Universit\`{a} di Napoli Federico II, Complesso Universitario di Monte Sant'Angelo, Via Cintia, 80126 Napoli, Italy}
\author {Alexandre Dauphin}
\email{Alexandre.dauphin@icfo.eu}
\affiliation{ICFO -- Institut de Ciencies Fot\'oniques, The Barcelona Institute of Science and Technology, 08860 Castelldefels (Barcelona), Spain}
\author {Pietro Massignan}
\affiliation{ICFO -- Institut de Ciencies Fot\'oniques, The Barcelona Institute of Science and Technology, 08860 Castelldefels (Barcelona), Spain}
\affiliation{Departament de F\'isica, Universitat Polit\`ecnica de Catalunya, Campus Nord B4-B5, 08034 Barcelona, Spain}


\begin{abstract}
In two, three and even four spatial dimensions, the transverse responses experienced by a charged particle on a lattice in a uniform magnetic field are fully controlled by topological invariants called Chern numbers, which characterize the energy bands of the underlying Hofstadter Hamiltonian. 
These remarkable features, solely arising from the magnetic translational symmetry, are captured by Diophantine equations which relate the fraction of occupied states, the magnetic flux and the Chern numbers of the system bands. 
Here we investigate the close analogy between the topological properties of Hofstadter Hamiltonians and the
diffraction figures resulting from optical gratings. 
In particular, we show that there is a one-to-one relation between the above mentioned Diophantine equation and the Bragg condition determining the far-field positions of the optical diffraction peaks.
As an interesting consequence of this mapping, we discuss how the robustness of diffraction figures to structural disorder in the grating is a direct analogue of the robustness of transverse conductance in the Quantum Hall effect.
\end{abstract}
\maketitle

\begin{figure*}[t]
\centering
\includegraphics[width=0.92\textwidth]{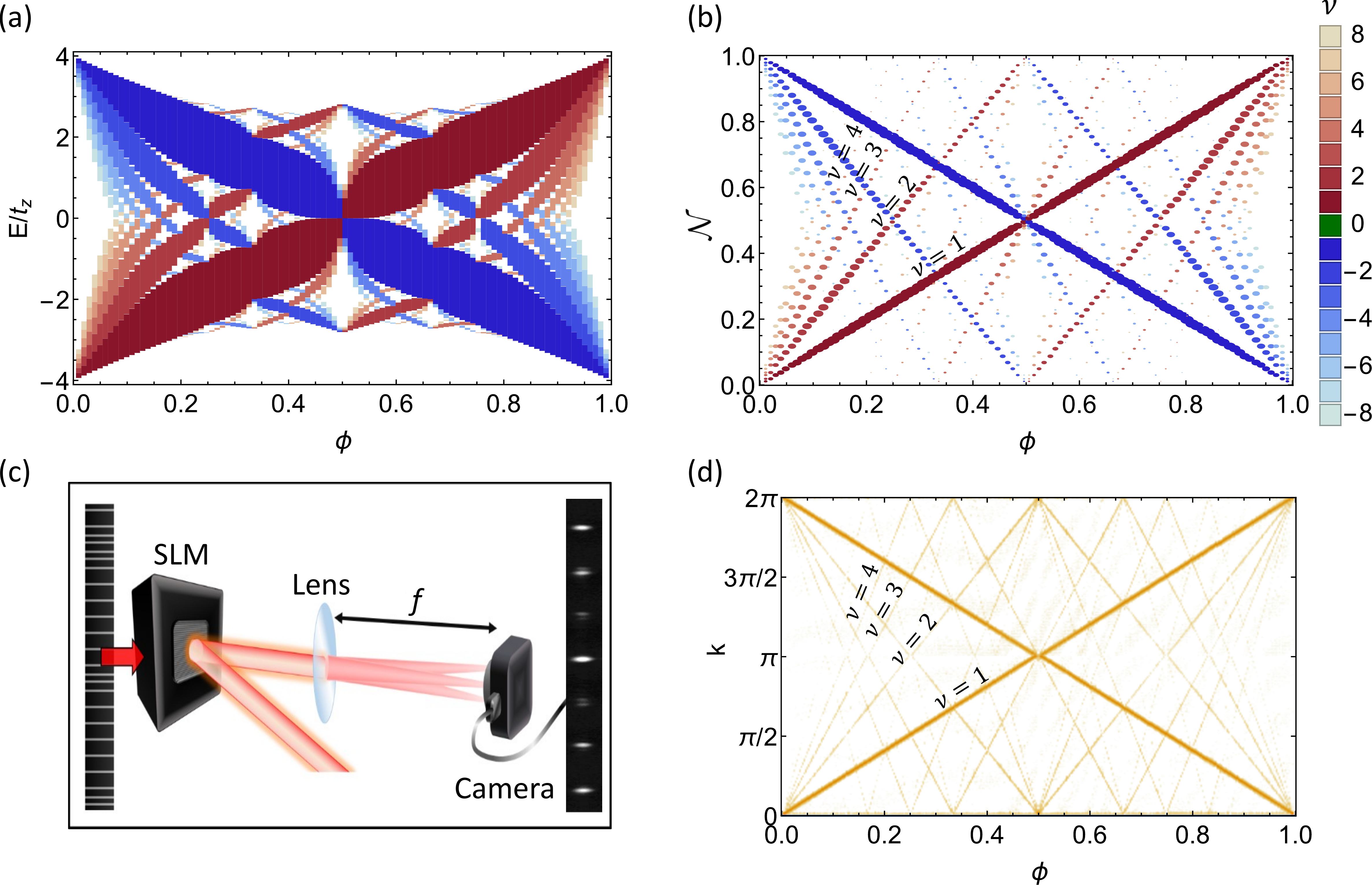}
\caption{\label{fig:fig1}
{\bf(a)} Energy spectrum of the 2D Hofstadter model, with equal hopping amplitudes $t_x=t_z$. The Hall conductivities of the gaps are encoded in the color scale.  {\bf(b)} The Wannier diagram shows the integrated density of states $\mathcal{N}$ below the gaps. The size of the dots is proportional to the amplitude of the gaps and their color is associated with the value of the Chern number $\nu$. The slopes of the lines in the diagram are directly given by the Streda-Widom formula, Eq.\ \eqref{eq:Diophantine}, and are independent of the hopping amplitudes.    
{\bf(c)} Experimental setup. A spatial light modulator (SLM) generates a grating with characteristic function $\chi_\phi(x) = \sign[\cos(2\pi \phi x)+d]$, whose diffraction figure is collected by a camera placed in the focal plane of a lens. \textbf{(d)} Experimental diffraction diagram, obtained by stacking a collection of diffraction figures for increasing values of the inverse spatial period $\phi$.}
\end{figure*}

Exposing an electronic crystal to a magnetic field radically alters its physical properties. A celebrated example is the integer quantum Hall effect (IQHE), where a perpendicular magnetic field converts a two-dimensional (2D) metal into a Chern insulator~\cite{Vonklitzing1986}. 
This physics is captured by the fractal spectrum (or ``butterfly") of the Hofstadter model~\cite{Hofstadter1976}, which features smooth bands characterized by integer topological invariants named Chern numbers. The transverse conductance in each energy gap, depicted in Fig.~\ref{fig:fig1}(a), is given by the sum of the Chern numbers of the occupied bands~\cite{TKNN1982,Kohmoto1989}, which ensures their robustness against local perturbations, such as interactions or disorder~\cite{niu_1985}. 
The physics becomes notably richer in higher spatial dimensions. A 3D metal exposed to a magnetic field was shown to exhibit quantized transverse conductivities in various 2D planes, determined by a triad of total first Chern numbers~\cite{Avron1983,Halperin1987,Kohmoto1990,Kohmoto1992,Haavasoja1984,Stormer1986,Druist1998,Koshino2001,Koshino2003,bruning2004,roy2016,lu2018topological}. In 4D, the response of the system to an external electro-magnetic field is governed by a more complex invariant called second Chern number~\cite{Avron1988,frohlich2000,zhang2001four, qi2008,edge2012metallic,Kraus2013,Price2015,Price2016,Price2018,Lohse2018,Zilberberg2018, Sugawa2018,wang2020circuit,chen2021,Petrides2020}.

Currently, very strong {\it synthetic} magnetic fields can be engineered in a variety of quantum simulators~\cite{Goldman2014,Ozawa2018,Cooper2018}, allowing for the observation of the 2D and 4D IQHE in various atomic~\cite{Aidelsburger2014,Lohse2018,Asteria19}, photonic~\cite{Kraus2012,Zilberberg2018,Mittal2019,DErrico20} and acoustic~\cite{Ni2019,chen2021} platforms. On the other hand, the 3D case was recently observed in a solid-state ZrTe$_5$ crystal \cite{Tang2019}. In general, traditional transport measurements in artificial platforms proved feasible~\cite{Krinner2014}, but not straightforward. As such, strong efforts are presently being undertaken to develop new methods for the characterization of the underlying topology~\cite{oktel2008,Kraus2012,Bardyn:2014,Tran17,wang17,Cardano2017,Asteria19,Tarnowski19}. 

In this work, we show that the topological features of the 2D, 3D and 4D IQHE can be linked to the properties of optical diffraction figures. Indeed, the peculiar translational symmetry of the Hofstadter Hamiltonian can be encoded in periodic phase gratings (diffraction gratings) dynamically generated by a spatial light modulator (SLM), and the magnetic flux piercing the lattice may be controlled by adjusting their spatial period. As we will show, the diffraction pattern produced by the grating bears close analogies to the {\it Wannier diagram} resulting from the Hofstadter model~\cite{Wannier1978}, and the slopes of the various lines directly yield the Chern numbers of the corresponding spectral gaps.

\section{Lorentz vs. density responses} 
The Chern numbers of the 2D IQHE can  be measured through the response of a system to (synthetic) electromagnetic fields. 

One class of such experiments measures the {\it Lorentz-type} response to a weak force. This is realized, for example, by preparing a localized wavefunction and reading out the displacement of its center-of-mass along the direction perpendicular to the force~\cite{Price2012,Dauphin2013,Aidelsburger2014,DErrico20,Price2016}, or by applying dimensional reduction to obtain an effective 1D time-periodic model and observing the consequent quantized displacement across the bulk of the system~(``Thouless pumping'')~\cite{Thouless1983,Lohse2015,Nakajima2016}. The latter approach has also been adopted for irrational values of the flux $\phi$, which give rise, through dimensional reduction, to 1D quasi-crystals~\cite{Shechtman1984}, i.e.~crystalline structures which are not periodic, but nonetheless exhibit long-range order~\cite{Kraus2012, Prodan2015, Dareau2017, Tanese2014, Zilberberg2021}. 
In particular, the Chern number of a 1D quasi-crystal has been measured in a diffraction experiment through a Lorentz-type measurement in Ref.~\cite{Dareau2017}. 

In a second line of experiments, which is the one we follow here, one instead probes a \textit{density-type} response by gradually changing the magnetic flux piercing the lattice.
As first discussed by Wannier in Ref.~\cite{Wannier1978}, even a small change in the flux radically modifies the underlying band structure, thereby altering the integrated density of states (i.e.\ the fraction of occupied bands). 
This may be conveniently displayed in a so-called {\it Wannier diagram}, which shows how the density of states grows linearly with the magnetic flux. Subsequent works~\cite{Streda1982, Streda1982bis,Dana1985} proved that the linear coefficient which links the density of states and the magnetic flux is proportional to the Hall conductivity. In particular, these studies highlighted that the complete topological information about the 2D Hofstadter model is fully encoded in its Wannier diagram~\cite{Dana1985}. 
As we will show in the following, this applies also to its generalizations to three~\cite{Kohmoto1990,Kohmoto1992} and four~\cite{Kraus2013,Price2015,Price2016} dimensions.

\section{2D Hofstadter model} 
\label{sec:2DHM}

Let us consider spinless fermions on a square lattice in the $xz$ plane, subject to a magnetic flux $\phi=p/q$ per plaquette, with integers $p$ and $q$ (we set $e=h=a=1$, where $a$ is the lattice spacing). In the Landau gauge with periodic boundary conditions in the $z$-direction, the eigenvalue problem reduces to the Harper equation~\cite{Harper1955}, which reads
\begin{multline}\label{eq:Harper2D}
\left[E_{n\bk} + 2 t_z \cos(2\pi \phi x +k_z)\right]u_{n\bk}(x)=\\
- t_x \left[e^{i k_x} u_{n\bk} (x+1)+ e^{-ik_x} u_{n\bk} (x-1)\right],
\end{multline}
where $u_{n\bk}(x)$ is the Bloch function for the $n^{\rm th}$ band, with quasi-momentum $\mathbf{k}$ in the reduced Brillouin zone (RBZ): $k_x \in [0,2\pi/q[$ and $k_z \in  [0,2\pi[$. 

The energy spectrum as a function of the magnetic flux $\phi$, shown in Fig.~\ref{fig:fig1}(a), is the famous Hofstadter butterfly, which displays an intriguing fractal structure. At zero temperature, the Hall conductivity $\sigma_r^{xz}$ in the $r^{\rm th}$ energy gap  equals the sum of the Chern numbers of the $r$ occupied bands~\cite{TKNN1982}
\beq
\nu=\sum_{n\leq r} \int_{RBZ} \Omega_n^{xz}d^2k/2\pi,
\eeq
defined in terms of the Berry curvature of the $n^{\rm th}$ band $\Omega_n^{xz}\!=\!i\left[\langle \partial_{k_x} u_{n\bk} \vert \partial_{k_z} u_{n\bk}\rangle -\langle \partial_{k_z} u_{n\bk} \vert \partial_{k_x} u_{n\bk}\rangle \right]$. 
Alternatively, the Hall conductivity can also be derived from the Streda-Widom formula $\nu= \partial_{\phi} \mathcal{N}$~\cite{Streda1982}, where $\mathcal{N}\equiv r/q$ denotes the integrated density of states below the $r^{\rm th}$ energy gap. This formula is very general, and holds true also for incompressible fractional Chern insulator phases~\cite{Repellin2020}, where the Hall conductivity is directly proportional to the many-body Chern number. For the 2D Hofstadter model, the Streda-Widom formula leads to the celebrated Diophantine equation 
\begin{equation}\label{eq:Diophantine}
r/q=\phi \nu + s,
\end{equation}
which has a unique integer solution $\nu$ for a set of integers $p,q,r,s$, assuming $p$ and $q$ are coprime, $0<r<q$ and $\vert \nu\vert <q/2$. As shown by Dana, Avron and Zak~\cite{Dana1985}, the Diophantine equation can also be derived from the properties of the magnetic translation operators. Their elegant proof, which we review in Appendix \ref{app:DiophantineEq2D}, only requires the periodicity of the wavefunction. The Diophantine equation can also be derived with the help of perturbation theory in the weak-coupling limit $t_x\ll t_z$~\cite{Kohmoto1989}. Remarkably, within this limit one can show that both the transverse conductivity and the quasi-momenta at which energy gaps open follow the \emph{same Diophantine equation}.

Here, we take advantage of the relation between the Bragg condition and the Diophantine equation to recover the Hofstadter butterfly through a 1D diffraction experiment. In fact, by exhibiting the same spatial periodic structure as the Harper onsite (cosine) potential, Bragg diffraction proves to be an ideal candidate to study key topological features of the Hofstadter model. The Lorentz-type measurement of Ref.~\cite{Dareau2017} required direct access to the complex phase of the diffracted wave, which was ingeniously extracted through an interferometric scheme. Here we discuss a conceptually different and much simpler density-type experiment, which only requires measuring the far-field intensity of light. A detailed description of our optical setup can be found in Appendix \ref{app:setup}.

In a first experiment, we write on a SLM a 1D diffraction grating generated by the characteristic function 
\begin{equation}\label{eq:carf}
\chi_\phi(x) = \sign[\cos(2\pi \phi x)+d],
\end{equation}
which generates a periodic figure with spatial period $1/\phi$.
The choice of the non-linear ``sign" function is mainly dictated by a practical convenience, as it enables one to observe many harmonics, thus allowing for a more accurate reconstruction of the diffraction diagram. Furthermore, it can be displayed on the SLM with high fidelity and exploiting the full resolution of the device. In principle, however, any other grating preserving the spatial periodicity of the original magnetic unit cell could be adopted.
The diffraction figure collected out of this structure contains a series of sharp Bragg peaks, as shown in Fig.~\ref{fig:fig1}(c). 
The relative intensity of the peaks can be controlled by tuning the dimensionless constant $d$. We set $d=0.25$ to ensure the best visibility of the main orders of diffraction. The position of the peaks can be derived from the Fourier transform of the grating (explicitly computed in Appendix \ref{app:Analytic2D}), and is simply given by the {\it Bragg condition}
 \begin{equation}
 k/(2\pi) = \phi \nu + s,
 \end{equation}
 with $\nu$ and $s$ integer numbers, and $k/2\pi=l/L$ is the ratio between an integer $l$ and the lattice size $L$. The latter condition requires that diffraction peaks are spaced by integer multiples of the RBZ. The Bragg condition is therefore a direct analogue of the Diophantine equation \eqref{eq:Diophantine}. Figure~\ref{fig:fig1}(d) depicts the \textit{diffraction diagram}, obtained by arranging side-by-side the diffraction figures generated by the grating for increasing values of $\phi$. 
The slopes of the lines forming the diagram correspond to the values of $\nu$, and the diffraction diagram is in a 1-to-1 correspondence with the Wannier diagram obtained from the density of states shown in Fig.~\ref{fig:fig1}(b). 
We point out that we are probing the topology of the bands which emerges from the traditional derivation, i.e. by means of the \textit{Peierls substitution}~\cite{Kohmoto1989}. Our approach would also apply to the Hofstadter butterfly spectrum resulting from the exact (numerical) calculation, since the topological Wannier diagram is preserved~\cite{Janecek2013}.

We have also investigated the effects of structural disorder or different characteristic functions $\chi$ on the diffraction diagram, and found that its main features are robust, as expected for topological properties. This will be presented in detail in Sec.~\ref{sec:disorder}.  

\begin{figure*}[t]
\centering
\includegraphics[width=0.92\textwidth]{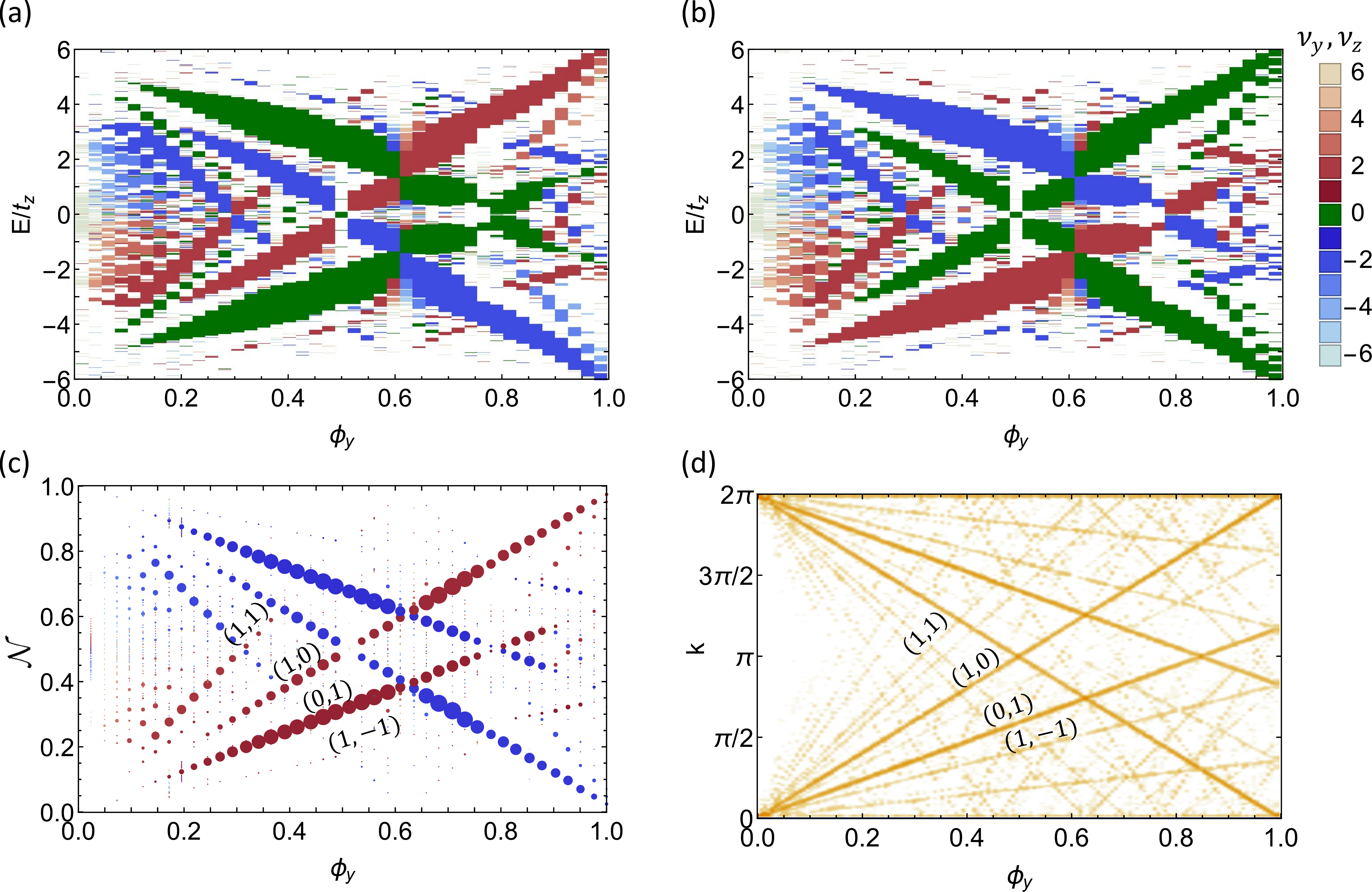}
\caption{\label{fig:fig2}
{\bf (a)}, {\bf (b)}  Energy spectrum of the 3D Hofstadter model. We choose hopping amplitudes $t_x=2t_y=2t_z$, and an approximately constant ratio between magnetic fluxes: $\phi_z/\phi_y \approx 3/5$, with $q_y=42$, $q_z=65$, and $p_y =p_z=P$. Panels (a) and (b) depict the values of the total Chern numbers $\nu_y$ and $\nu_z$, respectively, computed within each gap. {\bf (c)} Wannier diagram extracted from the integrated density of states. The sizes of the points are proportional to the amplitude of the corresponding gap and the color is associated with the value of $\Gamma$. 
{\bf (d)} Experimental diffraction diagram generated by $\psi(x)$, with $d=0.25$ and $\phi_z /\phi_y \approx 3/5$. We set $q_y=147$ and $q_z=250$, and let $p_y =p_z=P$ change from 0 to $q_y$.
}
\end{figure*}

\section{3D Hofstadter model}
The Hofstadter model is generalized to three spatial dimensions by considering a magnetic field having components along the three vectors spanning the cubic lattice~\cite{Kohmoto1990,Kohmoto1992,Koshino2001,Koshino2003,bruning2004}. Here we consider the particular case of a magnetic field $\textbf{B}=( 0,\phi_y ,\phi_z )$, 
where $\phi_{y}=p_{y}/q_{y}$ ($\phi_{z}=p_{z}/q_{z}$) is the magnetic flux through the $xz$ $(xy)$ plane. 
In the 3D analogue of the Landau gauge, the Harper equation reads
\begin{align}\label{eq:Harper3D}
&\left[E_{n\bk}+2 t_y \cos(2\pi \phi_z x +k_y) + 2t_z \cos(2\pi \phi_y x -k_z)\right]u_{n\bk}(x)
\nonumber \\&= - t_x \left[e^{i k_x } u_{n\bk} (x+1) + e^{-ik_x } u_{n\bk} (x-1)\right].
\end{align}
This eigenvalue problem has $Q$ solutions, $Q$ being the least common multiple of $q_y$ and $q_z$, and the RBZ is defined by $0<k_x\leq2\pi/Q$, $0<k_y,k_z\leq2\pi$. The separation among the $Q$ energy bands depends on the relative amplitudes of the hopping, and in the limit $t_x \gg t_y,t_z$ the system exhibits the maximum number of $Q-1$ gaps~\cite{Kohmoto1990}. These gaps are characterized by a triad of first Chern numbers $(\nu_x,\nu_y,\nu_z)$. These invariants appear in the quantization of the transverse conductivity in the planes defined by the unit vector of the cubic lattice. In our case, one finds $\sigma^{xy}=\nu_z$, $\sigma^{xz}=-\nu_y$ and $\sigma^{yz}=\nu_x=0$. The integrated density of states is related to the rational fluxes by a 3D version of the Diophantine equation~\cite{Kohmoto1990, Kohmoto1992}:
\begin{align}\label{eq:Dioph3d}
\mathcal{N} \equiv r/Q=  \phi_{y} \nu_y +\phi_z \nu_{z} +s,
\end{align}
where $s$ and $\nu_{\alpha}$ $(\alpha=\lbrace{y,z \rbrace})$ are integers. As in the 2D case, this relation arises directly from the magnetic translational symmetry of the system (see Appendix \ref{app:DiophantineEq3D}). The 3D Diophantine equation admits a unique solution for every set $r,p_y,p_z,q_y,q_z$, for $q_y$ and $q_z$ coprime integers, $0<r<Q$, $\vert \nu_y \vert < q_{y}/2$ and $\vert \nu_z \vert < q_{z}/2$. The transverse conductivity in the different planes  can therefore be computed with the help of the generalized Streda-Widom formula $\sigma_{ij}=  \epsilon_{ijk} \partial_{\phi_{k}} \mathcal{N}$. 

We now study the Hofstadter butterfly and the Wannier diagram resulting from the 3D quantum Hall lattice. There are many ways one can change the magnetic field, for example: i) changing its orientation with respect to the unit cell while keeping its amplitude fixed~\cite{Koshino2001}, ii) changing its amplitude while keeping its orientation fixed~\cite{Kohmoto1990,Koshino2003}. Here we adopt the second strategy, choosing the magnetic fluxes as $\phi_y = m_y P/Q$ and $\phi_z =m_z P/Q$. The Diophantine equation~\eqref{eq:Dioph3d} then reads 
\beq
\mathcal{N}= s+\Gamma P/Q,
\eeq
 where we have introduced a ``combined" first Chern number $\Gamma=\nu_{y} m_y + \nu_{z} m_z $.
Figures ~\ref{fig:fig2}(a) and \ref{fig:fig2}(b) show the Hofstadter butterfly generated for $t_x=2t_y=2t_z$ and a constant ratio between the fluxes $\phi_z/\phi_y\approx 3/5$. In the two panels, the color of the gaps corresponds to the values of $\nu_y$ and $\nu_z$, respectively. Figure~\ref{fig:fig2}(c) shows the Wannier diagram of the 3D Hofstadter model derived from the integrated density of states (here the colors of the points correspond to the values of $\Gamma$). As an optical topological probe, we arranged a second experiment where we write on the SLM a 1D diffraction grating generated by the characteristic function
\begin{equation}
\label{eq:carf3d}
\psi(x) = \sign[\cos(2\pi \phi_y x) +\cos(2\pi \phi_z x)+d]. 
\end{equation}
As in the 2D case, we set the constant $d=0.25$ to improve the visibility of the main diffraction orders.  The peaks are centered at $k=2\pi\left[\phi_y \nu_y + \phi_z \nu_z +s \right]$. Again, the translational symmetry of $\psi(x)$ ensures that the position of the Bragg peaks obeys the Diophantine equation~\eqref{eq:Dioph3d}. We reconstruct the Wannier diffraction diagram by arranging side-by-side the diffraction figures generated by the grating for different values of $\phi_y$. The slopes of the lines forming the diagram correspond to the values of $\Gamma$ [see Fig.~\ref{fig:fig2}(d)]. Also for the 3D case, we found an excellent agreement between the diffraction diagram and the Wannier diagram [see Fig.~\ref{fig:fig2}(c)].

\section{4D Hofstadter model}

\subsection{4D decoupled Hofstadter model}

The simplest model exhibiting the 4D IQHE is obtained by superposing two copies of the Hofstadter model, with a magnetic tensor having two non-zero components $B_{xz}= \phi_{x}$ and $B_{yw}=\phi_{y}$ \cite{Kraus2013,Prodan2015,Price2016}. The corresponding Hamiltonian leads to the following Harper equation:
\begin{align}\label{eq:Harper4D}
&E_{n\bk} u_{n\bk}(x,y) =\\ \nonumber 
&-\left[2t_z \cos(2\pi \phi_x x +k_z)+2t_w \cos(2\pi \phi_y y +k_w) \right]u_{n\bk}(x,y) \\ 
\nonumber &-t_{x} \left[e^{i k_x} u_{n \bk} (x+1,y) + e^{-ik_x} u_{n\bk} (x-1, y)\right]\\
\nonumber &-t_{y}\left[e^{i k_y} u_{n\bk} (x,y+1) + e^{-ik_y} u_{n \bk} (x, y-1)\right].
\end{align}
In this ``minimal" 4D model, as the planes having non-zero Berry curvatures are fully decoupled, the second Chern number $\mathcal{C}_2$ is simply the product of the two first Chern numbers $\nu_x$ and $\nu_y$. We will show that our approach also applies in the case in which the planes are coupled and the second Chern number is not factorizable.

\begin{figure*}[t]
\centering
\includegraphics[width=0.92\textwidth]{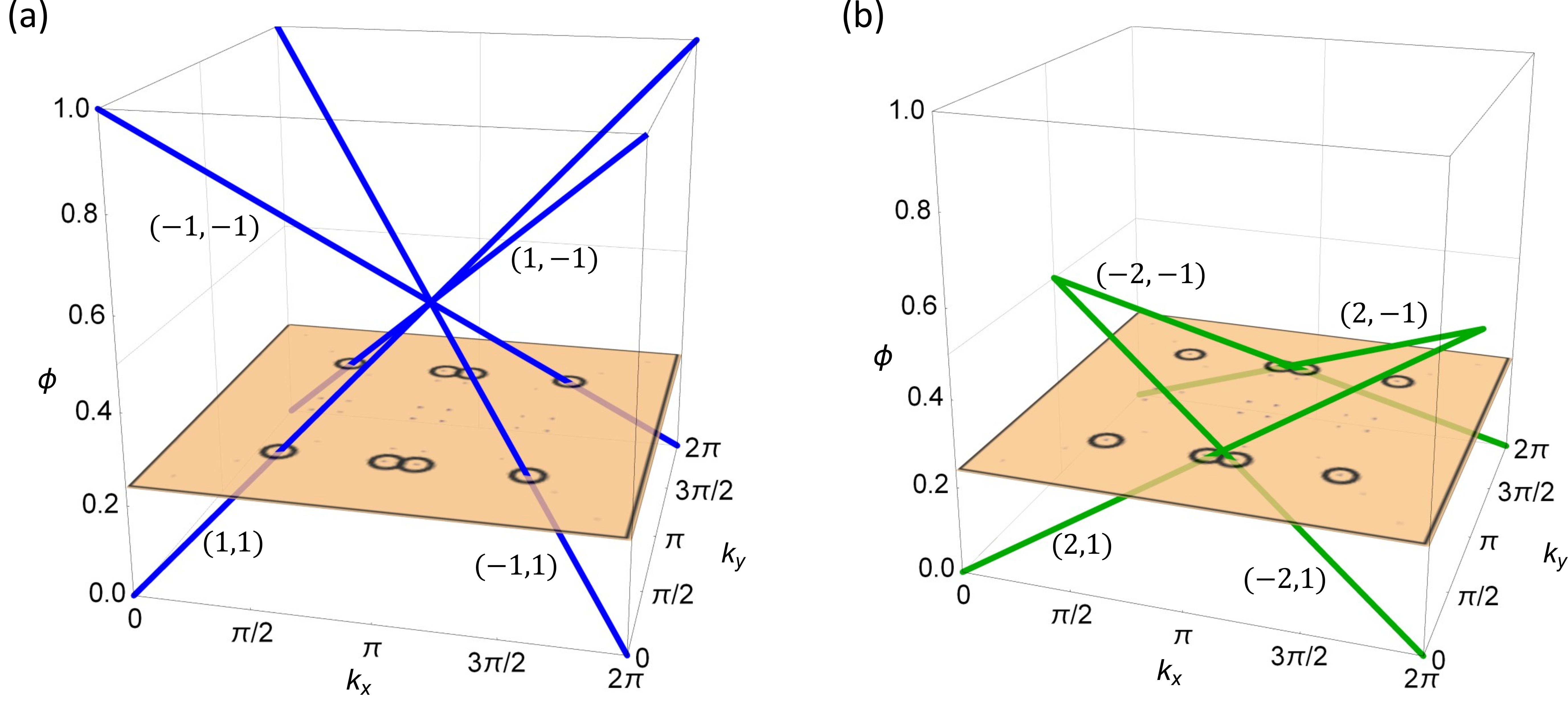}
\caption{\label{fig:fig3}
4D Hofstadter model. {\bf (a)},{\bf (b)} Expected diffraction diagram generated by $\xi(x,y)$ (simulation). The 3D spatial distribution of the diffraction peaks is characterized by multiple lines, each associated with different values of the first Chern numbers $\left(\nu_x,\nu_y \right)$. The particular case $\phi_x=\phi_y= \phi=7/30$ is depicted, where bright spots associated with different values of the first Chern numbers are enclosed by black circles. To improve readability,
peaks distributions for $\text{max} |{\lbrace \nu_x , \nu_y \rbrace } | = 1$ (blue lines in panel (a)) and $\text{max} |{\lbrace \nu_x , \nu_y \rbrace } | = 2$ (green lines in panel (b)) are plotted separately.
}
\end{figure*}

After invoking the two commuting magnetic translation symmetries of the system along $x$ and $y$, one is led to two independent Diophantine equations, which may be multiplied to obtain
\begin{equation}\label{eq:Dioph4D}
\mathcal{A}= ( \nu_x \phi_x +s_x) (\nu_y \phi_y +s_y),
\end{equation}
where $\mathcal{A}$ is the fraction of filled bands in the system at zero temperature.
As familiar from the {\it Gedanken-experiment} proposed by Laughlin~\cite{Laughlin1981}, changing the magnetic flux across the system effectively induces a ``density-response'' which alters the band-filling in a way that is proportional to the Chern numbers. In particular, the first total Chern numbers $\nu_x$ and $\nu_y$ are given by $\nu_{\alpha} =(1/s_{\beta}) \left.\partial_{\phi_{\alpha}}\mathcal{A}\right|_{\phi_{\beta}=0}$, and the second total Chern number is $\mathcal{C}_2 = \partial_{\phi_x} \partial_{\phi_y} \mathcal{A}= \nu_x\nu_y$. 
In a third experimental realization, we would suggest to write on the SLM a 2D diffraction grating generated by the characteristic function
\beq\label{charactFunctDecoupled}
\xi(x,y) = {\rm sign} \left[ \cos{(2\pi \phi x)} +  \cos{(2\pi \phi y)}+d\right]
,
\eeq
where we set $\phi_x=\phi_y=\phi$. The diffraction pattern generated by the grating defined above exhibits a series of bright spots placed at $\bk= 2\pi\left\lbrace (\phi \nu_{x} + s_{x}),(\phi \nu_{y} + s_{y}) \right\rbrace$ [see Fig.~\ref{fig:fig3}]. As in the previous experiments, a constant offset $d$ may be added to reinforce the visibility of the main spectral gaps. After stacking the 2D diffraction figures generated for different values of $\phi$ in a 3D plot, a Wannier diffraction diagram is obtained, where the coordinates of the points are given by two Diophantine equations, and the area under each line is given by Eq.~\eqref{eq:Dioph4D} [see Fig.~\ref{fig:fig3}]. We finally emphasise that this setup also allows one to reproduce diffraction figures featuring the same topology (in the sense discussed in Sec.~\ref{sec:2DHM}) as a class of 2D quasi-crystals, corresponding to the dimensional reduction of the 4D quantum Hall effect with irrational fluxes~\cite{Kraus2013,Lohse2018,Zilberberg2018}.

\subsection{4D coupled Hofstadter model}
\label{subsec:4Dcoupled}

Until now we focused on the simplest version of a 4D decoupled Hofstadter model, where the second Chern number could be factorized as a product of two first Chern numbers. More generally, let us consider the characteristic function
\begin{align} \label{FunctCoupled}
\Psi(x)={\rm sign}\{\cos(2\pi\phi_1 x)+\cos[2\pi \phi_2 (\epsilon x+y)]\},
\end{align} 
with $\epsilon$ being either 0 (yielding the decoupled model) or 1 (yielding the coupled model introduced in Ref.~\cite{Mochol-Grzelak2018}, where the factorization of the second Chern number no longer holds). This function is invariant under discrete translations, with crystal vectors ${\bf a}_1=q_1(1,-\epsilon)$ and ${\bf a}_2=q_2(0,1)$. The reciprocal vectors are therefore ${\bf b}_1=\frac{2\pi}{q_1}(1,0)$ and ${\bf b}_2=\frac{2\pi}{q_2}(\epsilon,1)$, so that ${\bf a}_i\cdot {\bf b}_j=2\pi \delta_{ij}$. The edges of the RBZ are quasi-momenta of the form
\beq
\bk=r_1 {\bf b}_1+r_2 {\bf b}_2,
\eeq
where $r_1=p_1 \nu_1+q_1 s_1$ and $r_2=p_2 \nu_2+q_2 s_2$.
This gives
\beq\label{peakLocation}
\frac{{\bk}}{2\pi}=\left(\frac{r_1}{q_1}+\epsilon\frac{r_2}{q_2},\frac{r_2}{q_2}\right).
\eeq
The second Chern number is given by
\begin{align}
\mathcal{C}_2=
\frac{\partial^2\left(\frac{r_1 r_2}{q_1q_2}\right)}{\partial\phi_1 \partial \phi_2} =
 \frac{\partial^2}{\partial\phi_1\partial \phi_2}
\left[\left(\frac{k_x}{2\pi}-\epsilon\frac{k_y}{2\pi}\right)\frac{k_y}{2\pi}
\right]\\ \nonumber
=\frac{\partial^2}{\partial\phi_1\partial \phi_2} \left(\frac{k_xk_y}{(2\pi)^2}-\epsilon\frac{k_y^2}{(2\pi)^2}\right).
\end{align}
In this more general framework, it is always possible to relate the second Chern number to the position of the diffraction peaks, even though the second Chern number does not factorize.

\section{Robustness against disorder}
\label{sec:disorder}

\begin{figure*}[!ht]
\includegraphics[width=\textwidth]{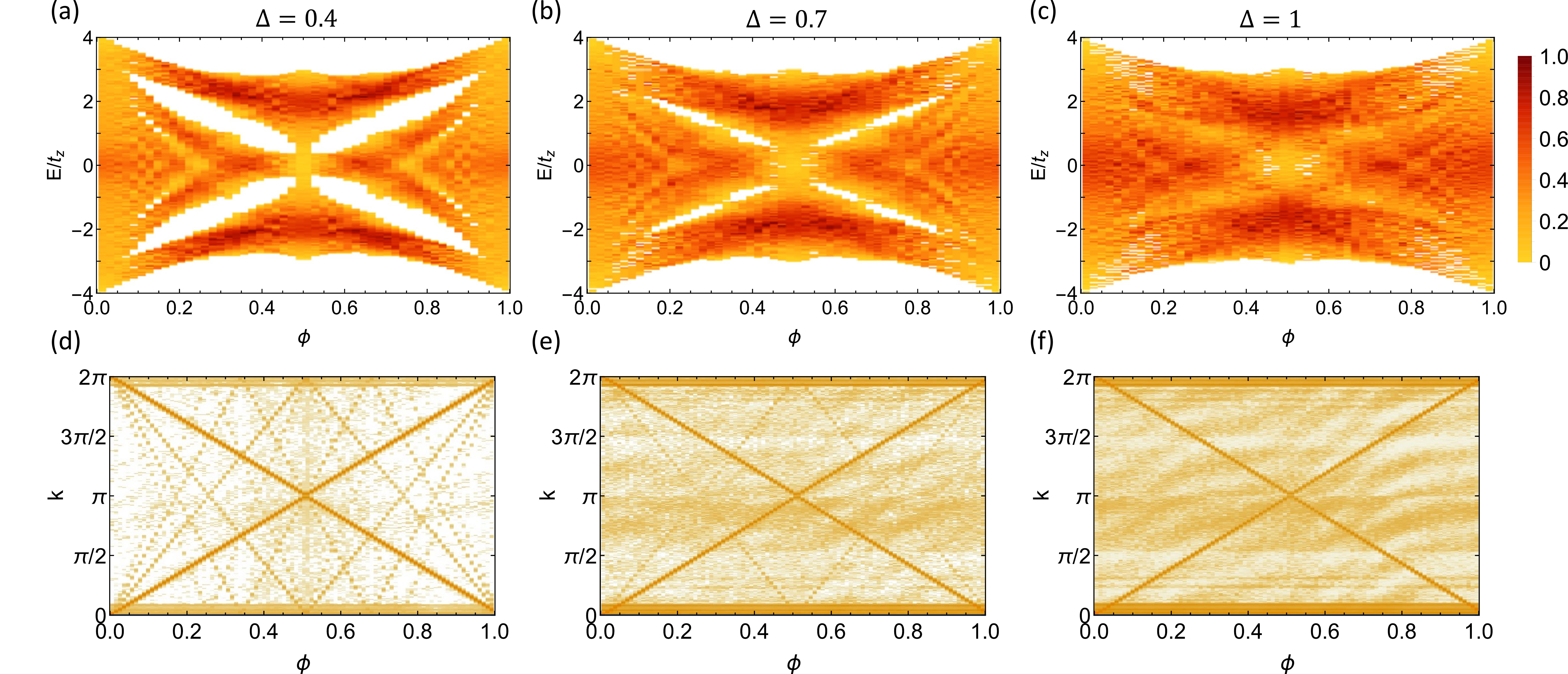}
\caption{\label{fig:fig4}
{\bf(a)-(c)} Hofstadter spectra in presence of spatial (on-site) disorder, with $t_x=t_z=1$. Energy values (in units of the hopping $t_z$) have been divided in bins of amplitude $\delta_{E}=0.05$, and each bin is coloured according to the fraction of states whose energy falls within the range $[E-\delta_E/2,E+\delta_E/2]$ (see Legend).
{\bf(d)-(f)} Experimental Wannier diffraction diagrams. Plots in each panel result from an average over 10 realizations with disorder. Different disorder strengths are probed: $\Delta=0.4$ (panels (a),(d)), $\Delta=0.7$ (panels (b),(e)), $\Delta=1$ (panels (c),(f)).
}
\end{figure*}

In presence of structural disorder, topological systems are known to be robust as long as the perturbation is much smaller than the corresponding spectral gap. Here we probe, both numerically and experimentally, the robustness of the Wannier diffraction diagram corresponding to the 2D Hofstadter model.\\
Figures~\ref{fig:fig4} (a)-(c) show the energy spectra of the 2D Hofstadter model, in presence of a disorder potential $V_{\rm dis} = t_z \, \xi(x)$, where $\xi(x)$ represents random numbers ranging in $[-\Delta,\Delta]$, $\Delta$ denoting the disorder strength. The on-site disorder has been optically implemented by slightly altering the 1D characteristic function: 
\begin{equation}\label{eq:carfdis}
\chi_{\rm dis}(x) = \sign[\cos(2\pi \phi x)+d
+\xi(x)],
\end{equation}
with $d=0.25$.

Experimental diagrams obtained for the different disorder strengths are plotted in Figs.~\ref{fig:fig4} (d)-(f). Each plot in Fig.~\ref{fig:fig4} results from an average over 10 realizations in presence of the picked level of disorder.
It is evident that diffraction orders associated with small spectral gaps are rapidly washed away, while the $1^{\text{st}}$-order Bragg peaks, which are associated with the largest gaps of the spectrum, remain visible even for the largest disorder (i.e.~$\Delta=1$). This proves that the main properties of the Wannier diffraction diagram are robust against disorder, as expected for topological features.

\section{Conclusions and Outlook}
In this paper, we showed that the topological properties of crystal electrons in presence of a constant magnetic field in 2, 3 and 4 spatial dimensions solely depend on the translational symmetry of the effective magnetic lattices. By means of a compact and intuitive genuinely-optical architecture, we succeeded in linking the topological invariants of the Hofstadter Hamiltonians to diffraction figures generated by photonic gratings that reflect the essential symmetries of the original model. Remarkably, such a connection is established within a single experimental realization, in contrast to other methods which typically rely on populating a subset of states. This is especially notable for 3D and 4D, where multiple topological invariants are required for each gap and where complete topological characterizations have not yet been achieved in experiments. In the future, it will be interesting to further generalize our method to probe even higher-dimensional extensions or other classes of topological models. In 4D, for example, a non-zero second Chern number may arise in systems featuring time-reversal symmetry, e.g. with spin-dependent gauge fields, in which all first Chern numbers vanish~\cite{zhang2001four,qi2008,Kraus2013,Price2018}. In our implementation, the role of spin may be played by the polarization degree of freedom of the light beam.

\section*{Acknowledgements}
We wish to thank E. Prodan for very insightful discussions. 
FDC, ADE, LM and FC acknowledge support from the EU (Horizon 2020 program, under ERC PHOSPhOR 694683).
MM acknowledges support from the Foundational Questions Institute Fund (Grant number FQXi-IAF19-05), and from the ANR Research Collaborative Project ``Qu-DICE" (ANR-PRC-CES47).
HMP is supported by the Royal Society via grants UF160112, RGF\textbackslash EA\textbackslash 180121 and RGF\textbackslash R1\textbackslash 180071.
AD and ML acknowledge support from ERC AdG NOQIA, State Research Agency AEI (Severo Ochoa Center of Excellence CEX2019-000910-S, Plan National FIDEUA PID2019-106901GB-I00/10.13039/501100011033, FPI, QUANTERA MAQS PCI2019-111828-2/10.13039/501100011033), Fundaci\'o Privada Cellex, Fundaci\'o Mir-Puig, Generalitat de Catalunya (AGAUR Grant 2017 SGR 1341, CERCA program, QuantumCAT U16-011424, co-funded by ERDF Operational Program of Catalonia 2014-2020), EU Horizon 2020 FET-OPEN OPTOLogic (Grant 899794), and the National Science Centre, Poland (Symfonia Grant 2016/20/W/ST4/00314), Marie Sklodowska-Curie grant STRETCH No 101029393.
AD further acknowledges the financial support from a fellowship granted by la Caixa Foundation (ID 100010434, fellowship code LCF/BQ/PR20/11770012).
PM acknowledges support by the Spanish MINECO (FIS2017-84114-C2-1-P), and EU FEDER Quantumcat. This research was supported in part by the National Science Foundation under Grant No. NSF PHY-1748958.

\appendix

\section{Details of the experimental setup}
\label{app:setup}

The experiment has been realized by studying the far-field diffraction patterns of a He-Ne laser beam (operating wavelength: $\lambda=632.8$ nm) impinging on a spatial light modulator (SLM). The SLM is a liquid-crystal (LC) device where the extraordinary refractive index has an inhomogeneous distribution that can be controlled and dynamically changed with a computer. This is achieved by means of an array of $ 1920 \times 1152 $ pixels (of side 9.2 $\mu $m), each one consisting of an electrode whose voltage independently tunes the local out-of-plane orientation of the LCs below it. This allows to manipulate the spatial phase distribution of the impinging beam, e.g. to display the desired transmission functions, as long as the beam is polarized along the extra-ordinary axis of the LCs.

Assuming a plane-wave incident wave-front, when imaging the diffracted light on the focal plane of a converging lens, one accesses the Fourier transform of the chosen transmission function. For instance, the transmission function $\chi(x)=\textrm{sign} \left[ \cos(2\pi p/q\, x)+d \right]$ was implemented by displaying the phase pattern $g(x)=\pi(3+\chi(x))/2 $. Since the input beam polarization can be not perfectly aligned with the extraordinary axis of the liquid crystals, we added to $g(x)$ the blazing function $2\pi y/ \Lambda \textrm{  mod }2 \pi$, with $\Lambda$ chosen to be small enough to spatially separate the light diffracted by the SLM from the $0^{\text{th}}$ (unmodulated) diffraction order in the Fourier plane (in our experiment, a convenient choice was $\Lambda=150$ pixels).

We displayed the pattern associated with $\chi(x)$ choosing $q=353$ and varying $p$ from $0$ to $q$. The corresponding lattice step size was chosen equal to $n_x=4$ pixels, thus building a grating consisting of 480 sites. The Fourier-transforming lens had focal length $f=30$ cm. These parameters were chosen to avoid overlap of diffracted light spots, and to image all the Brillouin zone on the camera sensor. 
In the experiment, the effective lattice length $L$, to be considered for simulating numerically the diffraction pattern, was determined by the transverse width of the beam. Assuming an input Gaussian beam of transverse intensity proportional to $\exp(-2r^2/w^2)$, we measured a beam waist $w=1.72\pm0.05$ mm. This implies that we were illuminating a portion of the SLM corresponding to $\approx 100$ lattice sites. Numerical simulations of the diffraction of a plane-wave impinging on a finite lattice of $L=100$ sites reproduces with good approximation the intensity patterns detected experimentally (see next section).

\section{Analytic expression of the diffraction figure for the 2D Hofstadter experiment}
\label{app:Analytic2D}

Let us consider a grating $S_L$ made up of $L$ sites, whose periodicity is modulated by the square pulse $\chi(x)$ defined in Eq.~\eqref{eq:carf}, and let us define $d\equiv\sin(\delta/2)$. The diffracted field results
\begin{align}\label{eq:diff}
f(k)&=\frac{1}{\sqrt{L}}\sum_{x=0}^{L-1}\chi(x) e^{i k x}\\ \nonumber
&=\frac{1}{\sqrt{L}}\sum_{x=0}^{L-1}\sign\left[\cos\left(\frac{2\pi x p }{q}\right)+\sin(\delta/2)\right] e^{i k x },
\end{align}
where we restrict the choice of $\delta$ to the range $[0,\pi[$. Since the system is finite, $k$ takes discrete values: $k\equiv 2\pi l/L$, with $l\in[-L/2,L/2[$.

In order to explicitly evaluate Eq.~\eqref{eq:diff}, we replace $\chi(x)$ with its Fourier series. The real ($A_n$) and the imaginary ($B_n$) part of the Fourier coefficients are given by
\begin{eqnarray}\label{eq:fcoeff}
A_n &\equiv& \frac{1}{T}\int_{0}^{T} dx \, \chi(x) \cos(n \Omega x), \nonumber \\
B_n &\equiv& \frac{1}{T} \int_{0}^{T} dx \, \chi(x) \sin(n \Omega x),
\end{eqnarray}
where $\Omega\equiv 2\pi/T= 2\pi p/q=2\pi \phi$ is the frequency of the square pulse (i.e., $1/\phi$ is its spatial period) and $n\in \mathbb{N}$.
Performing these integrals we obtain:
\begin{equation}\label{eq:fcoeff2}
A_0 = \frac{\delta}{\pi} ,\qquad
A_n 
=\frac{ 2(-1)^{n+1} \sin{ [n \frac{(\pi-\delta)}{2}]} }{\pi n} ,\qquad 
B_n = 0.
\end{equation}
The constant $\delta$ controls the number of visible lines in the Wannier diffraction diagram. In order to resolve a large number of peaks, we choose $\delta$ so that the visibility of the $6^{\rm th}$ diffraction peak is maximized. Since the first maximum of $A_6(\delta)$ occurs at $\delta \sim 0.5$, we pick $d=\sin(\delta/2)\sim0.25$. The Fourier expansion of $\chi(x)$ reads
\begin{equation}\label{eq:fseries}
\chi(x)=\sum_{n=0}^{\infty}A_n \cos{(2\pi\phi n x)}=\sum_{n=1}^{q} \tilde A_n \cos{(2\pi \phi n x)},
\end{equation}
where $\tilde A_n= \sum_{s=0}^\infty A_{n+sq}$.
To reach the last equality, we used the periodicity of the function to fold the Fourier spectrum over a range of $q$ harmonics.
It becomes clear now that each harmonic component of the pulse corresponds to one eigenvalue of the Harper equation~\eqref{eq:Harper2D} in the limit of vanishing $t_x\ll t_z$, multiplied by a coefficient $\tilde{A}_n$ which modulates relative intensities. We stress here that the previous limit is properly addressed in the context of adiabatic perturbation theory, as rigorously discussed in Ref.~\cite{Kohmoto1989}. Under those conditions, the gaps open at momenta $k_x$ given by a Diophantine equation \emph{identical} to the one determining the conductivity.  
Inserting Eq.~\eqref{eq:fseries} in Eq.~\eqref{eq:diff}, we find
\begin{equation}\label{eq:diff2}
f(k)= \sum_{x=0}^{L-1} \sum_{n=1}^{q} \frac{\tilde{A}_n}{2\sqrt{L}} (e^{i [k-n \Omega]x}+e^{i [k+n \Omega]x}).
\end{equation}

The two summations can be swapped (since they both converge to finite quantities), so we can first evaluate the one over $x$:
\begin{align}\label{eq:geom}
\zeta_\pm
\equiv\sum_{x=0}^{L-1} e^{i (k\pm n\Omega)x}
= \frac{1- e^{\pm i 2\pi \phi L}}{1- e^{i 2\pi (\frac{l}{L}\pm n \phi)}},
\end{align}
where we used $k= 2\pi l/L$.
Therefore, the explicit expression of the Fourier expansion of the field reads
\begin{align}\label{eq:fourierdiff}
f(l)= &f_{0}(l)+ 
\sum_{n=1}^{\infty}\frac{ 2(-1)^{n+1} \sin{ [n \frac{(\pi-\delta)}{2}]} }{2 \pi n \sqrt{L}} (\zeta_+ + \zeta_-) \\ \nonumber 
= &f_{0}(l)+\sum_{n=1}^{q-1}\frac{\tilde{A}_n}{2\sqrt{L}} (\zeta_+ + \zeta_-),
\end{align}
where $f_{0}(l)\equiv \frac{2 \delta}{\pi} \frac{ 1-i \sin{\left(2\pi \phi L   \right)}} {1- e^{i 2\pi l /L}}$.

For $L\rightarrow \infty$, each term in  Eq.~\eqref{eq:fourierdiff} is a sum of two Dirac delta-like contributions: $\zeta_\pm \propto \delta\left(\frac{l}{L}\pm n\phi\right)$, with $\phi=p/q$. For  finite $L$, there are two possible scenarios: $L$ may or may not be a multiple of $q$. 
When $L$ is a multiple of $q$, namely $L= z q$ (with integer $z$), we find the same result as in the infinite case and $f_{n}(l)\neq 0$ only when $l/L = \pm np/q $, that is $l=\pm z n p$. Since the main contribution to the diffracted field comes from the first harmonic $n=1$, the brightest peaks appear at $l=\pm z p$. 
When $L$ is not a multiple of $q$, we find non-zero diffracted field for many more values of $l$,
which contribute to creating the fractal spectrum typical of quasi-crystals. Nevertheless, the brightest peaks still appear at $l=\pm z p$, with $z=\text{Floor}[ L/q]$.

\section{Diophantine equation in 2D}
\label{app:DiophantineEq2D}

Here we demonstrate how the 2D Diophantine equation may be derived from the magnetic translational symmetry of the lattice Hamiltonian \cite{Dana1985,Blas2004}.
Let us consider spinless fermions on a square lattice under the action of a constant magnetic field with a magnetic flux $\phi=p/q$ per plaquette. In the Landau gauge, the Hamiltonian reads
\begin{equation}
H=\sum_{m,n}-t_xc^\dagger_{m+1,n}c_{m,n}-t_z e^{-2\pi i \phi m} c^\dagger_{m,n+1}c_{m,n}+\text{h.c. .}
\end{equation}
In presence of a non-zero magnetic flux, this Hamiltonian is no longer commuting with the lattice translation operator $T_z$. This motivates the introduction of the ``magnetic translation operators"
\begin{equation}
\begin{split}
M_x&=\sum_{m,n}c^\dagger_{m+1,n}c_{m,n},\\
M_z&=\sum_{m,n}e^{-2\pi i \phi m} c^\dagger_{m,n+1}c_{m,n}.
\end{split}
\end{equation}
These operators commute with the Hamiltonian, but not with each other: $M_x M_z=e^{-2\pi i \phi}M_z M_x$. However, $(M_x)^q$ commutes with $M_z$, so that $\{H,(M_x)^q,M_z\}$ form a ``complete set of commuting operators". 
Therefore, one has to consider a magnetic unit cell of at least $q$ sites in the $x$-direction in order to apply the Bloch theorem. Let us first focus on the ground band, which we assume to be non-degenerate. The eigenstates of the Hamiltonian within this band are Bloch states of the form~\cite{aidelsburger2015artifi}
\begin{equation}
\psi_{\bf k}(m,n)=e^{-i (k_x m+k_z n)}e^{2\pi i \phi m n} u_{\bf k}(m),
\end{equation}
where $k_x$ and $k_z$ are varying in a RBZ, defined by $0<k_x\leq2\pi/q$ and $0<k_z\leq2\pi$, $u_{k_x,k_z}(m+q)=u_{k_x,k_z}(m)$ and we omit the band index and take the lattice spacings equal to one for simplicity. Since $M_x$ and $M_z$ commute with the Hamiltonian, $\psi_\bk$ can be simultaneous eigenvectors of both $M_x$ and $M_z$ (i.e., $\psi_\bk,\,M_x\psi_\bk,$ and $M_z\psi_\bk$ all have the same energy $\epsilon_\bk$, and represent the same state, up to a phase factor).
Such Bloch states satisfy the relations
\begin{equation}
\begin{split}
&(M_x)^q \psi_{\bk}(m,n)= \psi_{\bk}(m+q,n)=e^{ik_x q} \psi_{\bk}(m,n),\\
&M_z \psi_{\bk}(m,n)=e^{-2\pi i \phi m} \psi_{\bk}(m,n+1)=e^{ik_z} \psi_{\bk}(m,n),
\end{split}
\end{equation}
as required for the Bloch theorem. 

This expression has still some residual degrees of freedom originating from the momentum-space periodicity of $u_\bk$. Indeed, $M_x$ is a standard translation operator, so that $u_{k_x+2\pi/q,k_z}=u_{k_x,k_z}$ (omitting the spatial indices). Along the other direction, instead, $M_z$ is not a standard translation operator, so that Bloch theorem does not apply and there is some extra freedom: $u_{k_x,k_z+2\pi} = e^{i\zeta(\bk)}u_{k_x,k_z}$.
This implies
\beq
e^{i\zeta(\bk)}u_{\bk}=u_{k_x,k_z+2\pi}=u_{k_x+2\pi/q,k_z+2\pi}=e^{i\zeta(k_x+2\pi/q,k_z)}u_{\bk},
\eeq
so that $\zeta(k_x+2\pi/q,k_z)=\zeta(\bk)+2\pi\C_1$, with $\C_1$ an integer number. The simplest function satisfying this condition is $\zeta(\bk)=k_x q\C_1$, and $2\pi\C_1$ is the phase acquired as the wavefunction is parallely-transported around the whole RBZ, and therefore $\C_1$ may be readily identified as the Chern number of the ground band.

There is yet an additional gauge freedom arising from the non-commutativity of the magnetic translation operators.
Since
\beq
M_z M_x \psi_\bk = e^{2\pi i \phi} M_x M_z \psi_\bk = e^{i (k_z+2\pi \phi)}M_x \psi_\bk
\eeq
and $M_z \psi_{k_x,k_z+2\pi\phi}=e^{i (k_z+2\pi \phi)}\psi_{k_x,k_z+2\pi\phi}$,
we find that $M_x \psi_\bk$ and $\psi_{k_x,k_z+2\pi\phi}$ must represent the same state, up to a phase factor:
\beq
M_x \psi_\bk = e^{i \eta(\bk)}\psi_{k_x,k_z+2\pi\phi}.
\eeq
This leads to
\begin{align}
e^{i \eta(\bk)}\psi_{k_x,k_z+2\pi\phi} &= M_x \psi_\bk \\\nonumber
&= M_x \psi_{k_x+2\pi/q,k_z}\\\nonumber
&= e^{i \eta(k_x+2\pi/q,k_z)}\psi_{k_x+2\pi/q,k_z+2\pi\phi}\\\nonumber
&= e^{i \eta(k_x+2\pi/q,k_z)}\psi_{k_x,k_z+2\pi\phi}.
\end{align}
Proceeding as above, one finds that the simplest functional form for the phase factor is $\eta(\bk)= k_x q S_1$, with $S_1$ an integer number. 

To derive the Diophantine equation, let us now consider a translation by a whole magnetic cell (along its extended direction), which gives
\begin{align}
(M_x)^q \psi_\bk&=e^{i k_x q^2 S_1}\psi_{k_x,k_z+2\pi p}
=e^{i k_x q(q S_1+p \C_1)}\psi_{k_x,k_z}\nonumber\\
&=e^{i k_x q}\psi_{k_x,k_z},
\end{align}
where in the last step we have used the fact that the translation by a whole magnetic cell generated by $(M_x)^q$ results in a simple phase factor $e^{i k_x q}$. This finally leads to the Diophantine equation for the ground band:
\beq\label{dioph_ground}
p\,\C_1+q S_1=1.
\eeq
This equation has a unique solution provided that $p$ and $q$ are coprime and $|C_1|< q/2$.
Considering now multiple bands, the same treatment as above may be performed for each band, obtaining $p\,\C_i+q S_i=1$.
Let us now recall that the $i^{\text{th}}$ Chern number is linked to the Hall conductivities of the neighboring gaps by $\C_i=\nu_i-\nu_{i-1}$, and let us assume similarly that $S_i=s_i-s_{i-1}$, with $\nu_0=s_0=0$. 
Summing the equations over all the gaps, one readily obtains the usual Diophantine equation
\beq
p\nu_r+q s_r=r,
\eeq
where $\nu_r=\sum_{i=1}^r\C_i$ is the Hall conductivity. Again, the solution is unique provided that $|\nu_r|<q/2$.\\

\section{Diophantine equation in 3D}
\label{app:DiophantineEq3D}

By closely following the steps of the previous section, here we derive the 3D Diophantine equation Eq.~\eqref{eq:Dioph3d} from the magnetic translation symmetry of the 3D Hofstadter model~\cite{Blas2004}. 

First, we set the following periodic boundary conditions on the Bloch functions:
\begin{align}\label{eq:pbc}
u_{k_x + 2\pi/Q, k_y ,k_z}&=u_{\bk},\\ \nonumber
u_{k_x, k_y + 2\pi,k_z}&= e^{i k_x \sigma_y Q }u_{\bk},\\ \nonumber
u_{k_x, k_y ,k_z + 2\pi}&= e^{i k_x \sigma_z Q }u_{\bk},
\end{align}
where we refer to the band Chern numbers as $\sigma_{\alpha}$, and we omit the band index and set the lattice spacings equal to one again.

Applying the 3D magnetic translation operator $M_{x}$ to $u_{\bk}$, we obtain
\begin{align}
M_{x} u_{\bk}= e^{i S k_x Q}u_{k_x, k_y +2\pi \phi_y, k_z +2\pi \phi_z}.
\end{align}
By reiterating this application $Q$ times and invoking the boundary conditions~\eqref{eq:pbc}, one is led to 
\begin{align}\label{eq:proofDana1}
e^{i k_x Q}u_{\bk}=e^{ik_x Q(\sigma_y m_y P + \sigma_z m_z P + S Q)} u_{\bk},
\end{align}
where we set $\phi_y = m_y P/Q$ and $\phi_z = m_z P/Q$.
Equation~\eqref{eq:proofDana1} is satisfied if
\begin{equation}\label{eq:proofDana2}
\sigma_y  m_y P + \sigma_z m_z P + S Q =1.
\end{equation}
The 3D Diophantine equation
\begin{equation}
\nu^{(r)}_y  m_y P+ \nu^{(r)}_z m_z P + s^{(r)} Q =r
\end{equation}  
is obtained from Eq.~\eqref{eq:proofDana2}, by recalling $\sigma^{(r)}_{\alpha}= \nu^{(r)}_{\alpha}-\nu^{(r-1)}_{\alpha} $.

%



\begin{thebibliography}{70}%
\makeatletter
\providecommand \@ifxundefined [1]{%
 \@ifx{#1\undefined}
}%
\providecommand \@ifnum [1]{%
 \ifnum #1\expandafter \@firstoftwo
 \else \expandafter \@secondoftwo
 \fi
}%
\providecommand \@ifx [1]{%
 \ifx #1\expandafter \@firstoftwo
 \else \expandafter \@secondoftwo
 \fi
}%
\providecommand \natexlab [1]{#1}%
\providecommand \enquote  [1]{``#1''}%
\providecommand \bibnamefont  [1]{#1}%
\providecommand \bibfnamefont [1]{#1}%
\providecommand \citenamefont [1]{#1}%
\providecommand \href@noop [0]{\@secondoftwo}%
\providecommand \href [0]{\begingroup \@sanitize@url \@href}%
\providecommand \@href[1]{\@@startlink{#1}\@@href}%
\providecommand \@@href[1]{\endgroup#1\@@endlink}%
\providecommand \@sanitize@url [0]{\catcode `\\12\catcode `\$12\catcode
  `\&12\catcode `\#12\catcode `\^12\catcode `\_12\catcode `\%12\relax}%
\providecommand \@@startlink[1]{}%
\providecommand \@@endlink[0]{}%
\providecommand \url  [0]{\begingroup\@sanitize@url \@url }%
\providecommand \@url [1]{\endgroup\@href {#1}{\urlprefix }}%
\providecommand \urlprefix  [0]{URL }%
\providecommand \Eprint [0]{\href }%
\providecommand \doibase [0]{http://dx.doi.org/}%
\providecommand \selectlanguage [0]{\@gobble}%
\providecommand \bibinfo  [0]{\@secondoftwo}%
\providecommand \bibfield  [0]{\@secondoftwo}%
\providecommand \translation [1]{[#1]}%
\providecommand \BibitemOpen [0]{}%
\providecommand \bibitemStop [0]{}%
\providecommand \bibitemNoStop [0]{.\EOS\space}%
\providecommand \EOS [0]{\spacefactor3000\relax}%
\providecommand \BibitemShut  [1]{\csname bibitem#1\endcsname}%
\let\auto@bib@innerbib\@empty
\bibitem [{\citenamefont {von Klitzing}(1986)}]{Vonklitzing1986}%
  \BibitemOpen
  \bibfield  {author} {\bibinfo {author} {\bibfnamefont {K.}~\bibnamefont {von
  Klitzing}},\ }\bibfield  {title} {\bibinfo {title} {\emph {The quantized Hall
  effect}},\ }\href {\doibase 10.1103/RevModPhys.58.519} {\bibfield  {journal}
  {\bibinfo  {journal} {Rev. Mod. Phys.}\ }\textbf {\bibinfo {volume} {58}},\
  \bibinfo {pages} {519} (\bibinfo {year} {1986})}\BibitemShut {NoStop}%
\bibitem [{\citenamefont {Hofstadter}(1976)}]{Hofstadter1976}%
  \BibitemOpen
  \bibfield  {author} {\bibinfo {author} {\bibfnamefont {D.~R.}\ \bibnamefont
  {Hofstadter}},\ }\bibfield  {title} {\bibinfo {title} {\emph {Energy levels
  and wave functions of Bloch electrons in rational and irrational magnetic
  fields}},\ }\href {\doibase 10.1103/PhysRevB.14.2239} {\bibfield  {journal}
  {\bibinfo  {journal} {Phys. Rev. B}\ }\textbf {\bibinfo {volume} {14}},\
  \bibinfo {pages} {2239} (\bibinfo {year} {1976})}\BibitemShut {NoStop}%
\bibitem [{\citenamefont {Thouless}\ \emph {et~al.}(1982)\citenamefont
  {Thouless}, \citenamefont {Kohmoto}, \citenamefont {Nightingale},\ and\
  \citenamefont {den Nijs}}]{TKNN1982}%
  \BibitemOpen
  \bibfield  {author} {\bibinfo {author} {\bibfnamefont {D.~J.}\ \bibnamefont
  {Thouless}}, \bibinfo {author} {\bibfnamefont {M.}~\bibnamefont {Kohmoto}},
  \bibinfo {author} {\bibfnamefont {M.~P.}\ \bibnamefont {Nightingale}}, \ and\
  \bibinfo {author} {\bibfnamefont {M.}~\bibnamefont {den Nijs}},\ }\bibfield
  {title} {\bibinfo {title} {\emph {Quantized Hall Conductance in a
  Two-Dimensional Periodic Potential}},\ }\href {\doibase
  10.1103/PhysRevLett.49.405} {\bibfield  {journal} {\bibinfo  {journal} {Phys.
  Rev. Lett.}\ }\textbf {\bibinfo {volume} {49}},\ \bibinfo {pages} {405}
  (\bibinfo {year} {1982})}\BibitemShut {NoStop}%
\bibitem [{\citenamefont {Kohmoto}(1989)}]{Kohmoto1989}%
  \BibitemOpen
  \bibfield  {author} {\bibinfo {author} {\bibfnamefont {M.}~\bibnamefont
  {Kohmoto}},\ }\bibfield  {title} {\bibinfo {title} {\emph {Zero modes and the
  quantized Hall conductance of the two-dimensional lattice in a magnetic
  field}},\ }\href {\doibase 10.1103/PhysRevB.39.11943} {\bibfield  {journal}
  {\bibinfo  {journal} {Phys. Rev. B}\ }\textbf {\bibinfo {volume} {39}},\
  \bibinfo {pages} {11943} (\bibinfo {year} {1989})}\BibitemShut {NoStop}%
\bibitem [{\citenamefont {Niu}\ \emph {et~al.}(1985)\citenamefont {Niu},
  \citenamefont {Thouless},\ and\ \citenamefont {Wu}}]{niu_1985}%
  \BibitemOpen
  \bibfield  {author} {\bibinfo {author} {\bibfnamefont {Q.}~\bibnamefont
  {Niu}}, \bibinfo {author} {\bibfnamefont {D.~J.}\ \bibnamefont {Thouless}}, \
  and\ \bibinfo {author} {\bibfnamefont {Y.-S.}\ \bibnamefont {Wu}},\
  }\bibfield  {title} {\bibinfo {title} {\emph {Quantized Hall conductance as a
  topological invariant}},\ }\href {\doibase 10.1103/PhysRevB.31.3372}
  {\bibfield  {journal} {\bibinfo  {journal} {Phys. Rev. B}\ }\textbf {\bibinfo
  {volume} {31}},\ \bibinfo {pages} {3372} (\bibinfo {year}
  {1985})}\BibitemShut {NoStop}%
\bibitem [{\citenamefont {Avron}\ \emph {et~al.}(1983)\citenamefont {Avron},
  \citenamefont {Seiler},\ and\ \citenamefont {Simon}}]{Avron1983}%
  \BibitemOpen
  \bibfield  {author} {\bibinfo {author} {\bibfnamefont {J.~E.}\ \bibnamefont
  {Avron}}, \bibinfo {author} {\bibfnamefont {R.}~\bibnamefont {Seiler}}, \
  and\ \bibinfo {author} {\bibfnamefont {B.}~\bibnamefont {Simon}},\ }\bibfield
   {title} {\bibinfo {title} {\emph {Homotopy and Quantization in Condensed
  Matter Physics}},\ }\href {\doibase 10.1103/PhysRevLett.51.51} {\bibfield
  {journal} {\bibinfo  {journal} {Phys. Rev. Lett.}\ }\textbf {\bibinfo
  {volume} {51}},\ \bibinfo {pages} {51} (\bibinfo {year} {1983})}\BibitemShut
  {NoStop}%
\bibitem [{\citenamefont {Halperin}(1987)}]{Halperin1987}%
  \BibitemOpen
  \bibfield  {author} {\bibinfo {author} {\bibfnamefont {B.~I.}\ \bibnamefont
  {Halperin}},\ }\bibfield  {title} {\bibinfo {title} {\emph {Possible States
  for a Three-Dimensional Electron Gas in a Strong Magnetic Field}},\ }\href
  {\doibase 10.7567/jjaps.26s3.1913} {\bibfield  {journal} {\bibinfo  {journal}
  {Jpn. J. Appl. Phys.}\ }\textbf {\bibinfo {volume} {26}},\ \bibinfo {pages}
  {1913} (\bibinfo {year} {1987})}\BibitemShut {NoStop}%
\bibitem [{\citenamefont {Montambaux}\ and\ \citenamefont
  {Kohmoto}(1990)}]{Kohmoto1990}%
  \BibitemOpen
  \bibfield  {author} {\bibinfo {author} {\bibfnamefont {G.}~\bibnamefont
  {Montambaux}}\ and\ \bibinfo {author} {\bibfnamefont {M.}~\bibnamefont
  {Kohmoto}},\ }\bibfield  {title} {\bibinfo {title} {\emph {Quantized Hall
  effect in three dimensions}},\ }\href {\doibase 10.1103/PhysRevB.41.11417}
  {\bibfield  {journal} {\bibinfo  {journal} {Phys. Rev. B}\ }\textbf {\bibinfo
  {volume} {41}},\ \bibinfo {pages} {11417} (\bibinfo {year}
  {1990})}\BibitemShut {NoStop}%
\bibitem [{\citenamefont {Kohmoto}\ \emph {et~al.}(1992)\citenamefont
  {Kohmoto}, \citenamefont {Halperin},\ and\ \citenamefont {Wu}}]{Kohmoto1992}%
  \BibitemOpen
  \bibfield  {author} {\bibinfo {author} {\bibfnamefont {M.}~\bibnamefont
  {Kohmoto}}, \bibinfo {author} {\bibfnamefont {B.~I.}\ \bibnamefont
  {Halperin}}, \ and\ \bibinfo {author} {\bibfnamefont {Y.-S.}\ \bibnamefont
  {Wu}},\ }\bibfield  {title} {\bibinfo {title} {\emph {Diophantine equation
  for the three-dimensional quantum Hall effect}},\ }\href {\doibase
  10.1103/PhysRevB.45.13488} {\bibfield  {journal} {\bibinfo  {journal} {Phys.
  Rev. B}\ }\textbf {\bibinfo {volume} {45}},\ \bibinfo {pages} {13488}
  (\bibinfo {year} {1992})}\BibitemShut {NoStop}%
\bibitem [{\citenamefont {Haavasoja}\ \emph {et~al.}(1984)\citenamefont
  {Haavasoja}, \citenamefont {St\"ormer}, \citenamefont {Bishop}, \citenamefont
  {Narayanamurti}, \citenamefont {Gossard},\ and\ \citenamefont
  {Wiegmann}}]{Haavasoja1984}%
  \BibitemOpen
  \bibfield  {author} {\bibinfo {author} {\bibfnamefont {T.}~\bibnamefont
  {Haavasoja}}, \bibinfo {author} {\bibfnamefont {H.}~\bibnamefont
  {St\"ormer}}, \bibinfo {author} {\bibfnamefont {D.}~\bibnamefont {Bishop}},
  \bibinfo {author} {\bibfnamefont {V.}~\bibnamefont {Narayanamurti}}, \bibinfo
  {author} {\bibfnamefont {A.}~\bibnamefont {Gossard}}, \ and\ \bibinfo
  {author} {\bibfnamefont {W.}~\bibnamefont {Wiegmann}},\ }\bibfield  {title}
  {\bibinfo {title} {\emph {Magnetization measurements on a two-dimensional
  electron system}},\ }\href {\doibase
  https://doi.org/10.1016/0039-6028(84)90325-X} {\bibfield  {journal} {\bibinfo
   {journal} {Surf. Sci.}\ }\textbf {\bibinfo {volume} {142}},\ \bibinfo
  {pages} {294} (\bibinfo {year} {1984})}\BibitemShut {NoStop}%
\bibitem [{\citenamefont {St\"ormer}\ \emph {et~al.}(1986)\citenamefont
  {St\"ormer}, \citenamefont {Eisenstein}, \citenamefont {Gossard},
  \citenamefont {Wiegmann},\ and\ \citenamefont {Baldwin}}]{Stormer1986}%
  \BibitemOpen
  \bibfield  {author} {\bibinfo {author} {\bibfnamefont {H.~L.}\ \bibnamefont
  {St\"ormer}}, \bibinfo {author} {\bibfnamefont {J.~P.}\ \bibnamefont
  {Eisenstein}}, \bibinfo {author} {\bibfnamefont {A.~C.}\ \bibnamefont
  {Gossard}}, \bibinfo {author} {\bibfnamefont {W.}~\bibnamefont {Wiegmann}}, \
  and\ \bibinfo {author} {\bibfnamefont {K.}~\bibnamefont {Baldwin}},\
  }\bibfield  {title} {\bibinfo {title} {\emph {Quantization of the Hall effect
  in an anisotropic three-dimensional electronic system}},\ }\href {\doibase
  10.1103/PhysRevLett.56.85} {\bibfield  {journal} {\bibinfo  {journal} {Phys.
  Rev. Lett.}\ }\textbf {\bibinfo {volume} {56}},\ \bibinfo {pages} {85}
  (\bibinfo {year} {1986})}\BibitemShut {NoStop}%
\bibitem [{\citenamefont {Druist}\ \emph {et~al.}(1998)\citenamefont {Druist},
  \citenamefont {Turley}, \citenamefont {Maranowski}, \citenamefont {Gwinn},\
  and\ \citenamefont {Gossard}}]{Druist1998}%
  \BibitemOpen
  \bibfield  {author} {\bibinfo {author} {\bibfnamefont {D.~P.}\ \bibnamefont
  {Druist}}, \bibinfo {author} {\bibfnamefont {P.~J.}\ \bibnamefont {Turley}},
  \bibinfo {author} {\bibfnamefont {K.~D.}\ \bibnamefont {Maranowski}},
  \bibinfo {author} {\bibfnamefont {E.~G.}\ \bibnamefont {Gwinn}}, \ and\
  \bibinfo {author} {\bibfnamefont {A.~C.}\ \bibnamefont {Gossard}},\
  }\bibfield  {title} {\bibinfo {title} {\emph {Observation of Chiral Surface
  States in the Integer Quantum Hall Effect}},\ }\href {\doibase
  10.1103/PhysRevLett.80.365} {\bibfield  {journal} {\bibinfo  {journal} {Phys.
  Rev. Lett.}\ }\textbf {\bibinfo {volume} {80}},\ \bibinfo {pages} {365}
  (\bibinfo {year} {1998})}\BibitemShut {NoStop}%
\bibitem [{\citenamefont {Koshino}\ \emph {et~al.}(2001)\citenamefont
  {Koshino}, \citenamefont {Aoki}, \citenamefont {Kuroki}, \citenamefont
  {Kagoshima},\ and\ \citenamefont {Osada}}]{Koshino2001}%
  \BibitemOpen
  \bibfield  {author} {\bibinfo {author} {\bibfnamefont {M.}~\bibnamefont
  {Koshino}}, \bibinfo {author} {\bibfnamefont {H.}~\bibnamefont {Aoki}},
  \bibinfo {author} {\bibfnamefont {K.}~\bibnamefont {Kuroki}}, \bibinfo
  {author} {\bibfnamefont {S.}~\bibnamefont {Kagoshima}}, \ and\ \bibinfo
  {author} {\bibfnamefont {T.}~\bibnamefont {Osada}},\ }\bibfield  {title}
  {\bibinfo {title} {\emph {Hofstadter Butterfly and Integer Quantum Hall
  Effect in Three Dimensions}},\ }\href {\doibase 10.1103/PhysRevLett.86.1062}
  {\bibfield  {journal} {\bibinfo  {journal} {Phys. Rev. Lett.}\ }\textbf
  {\bibinfo {volume} {86}},\ \bibinfo {pages} {1062} (\bibinfo {year}
  {2001})}\BibitemShut {NoStop}%
\bibitem [{\citenamefont {Koshino}\ and\ \citenamefont
  {Aoki}(2003)}]{Koshino2003}%
  \BibitemOpen
  \bibfield  {author} {\bibinfo {author} {\bibfnamefont {M.}~\bibnamefont
  {Koshino}}\ and\ \bibinfo {author} {\bibfnamefont {H.}~\bibnamefont {Aoki}},\
  }\bibfield  {title} {\bibinfo {title} {\emph {Integer quantum Hall effect in
  isotropic three-dimensional crystals}},\ }\href {\doibase
  10.1103/PhysRevB.67.195336} {\bibfield  {journal} {\bibinfo  {journal} {Phys.
  Rev. B}\ }\textbf {\bibinfo {volume} {67}},\ \bibinfo {pages} {195336}
  (\bibinfo {year} {2003})}\BibitemShut {NoStop}%
\bibitem [{\citenamefont {Br\"uning}\ \emph {et~al.}(2004)\citenamefont
  {Br\"uning}, \citenamefont {Demidov},\ and\ \citenamefont
  {Geyler}}]{bruning2004}%
  \BibitemOpen
  \bibfield  {author} {\bibinfo {author} {\bibfnamefont {J.}~\bibnamefont
  {Br\"uning}}, \bibinfo {author} {\bibfnamefont {V.~V.}\ \bibnamefont
  {Demidov}}, \ and\ \bibinfo {author} {\bibfnamefont {V.~A.}\ \bibnamefont
  {Geyler}},\ }\bibfield  {title} {\bibinfo {title} {\emph {Hofstadter-type
  spectral diagrams for the Bloch electron in three dimensions}},\ }\href
  {\doibase 10.1103/PhysRevB.69.033202} {\bibfield  {journal} {\bibinfo
  {journal} {Phys. Rev. B}\ }\textbf {\bibinfo {volume} {69}},\ \bibinfo
  {pages} {033202} (\bibinfo {year} {2004})}\BibitemShut {NoStop}%
\bibitem [{\citenamefont {Roy}\ \emph {et~al.}(2016)\citenamefont {Roy},
  \citenamefont {Kolodrubetz}, \citenamefont {Moore},\ and\ \citenamefont
  {Grushin}}]{roy2016}%
  \BibitemOpen
  \bibfield  {author} {\bibinfo {author} {\bibfnamefont {S.}~\bibnamefont
  {Roy}}, \bibinfo {author} {\bibfnamefont {M.}~\bibnamefont {Kolodrubetz}},
  \bibinfo {author} {\bibfnamefont {J.~E.}\ \bibnamefont {Moore}}, \ and\
  \bibinfo {author} {\bibfnamefont {A.~G.}\ \bibnamefont {Grushin}},\
  }\bibfield  {title} {\bibinfo {title} {\emph {Chern numbers and chiral
  anomalies in Weyl butterflies}},\ }\href {\doibase
  10.1103/PhysRevB.94.161107} {\bibfield  {journal} {\bibinfo  {journal} {Phys.
  Rev. B}\ }\textbf {\bibinfo {volume} {94}},\ \bibinfo {pages} {161107}
  (\bibinfo {year} {2016})}\BibitemShut {NoStop}%
\bibitem [{\citenamefont {Lu}\ \emph {et~al.}(2018)\citenamefont {Lu},
  \citenamefont {Gao},\ and\ \citenamefont {Wang}}]{lu2018topological}%
  \BibitemOpen
  \bibfield  {author} {\bibinfo {author} {\bibfnamefont {L.}~\bibnamefont
  {Lu}}, \bibinfo {author} {\bibfnamefont {H.}~\bibnamefont {Gao}}, \ and\
  \bibinfo {author} {\bibfnamefont {Z.}~\bibnamefont {Wang}},\ }\bibfield
  {title} {\bibinfo {title} {\emph {{Topological one-way fiber of second Chern
  number}}},\ }\href {\doibase 10.1038/s41467-018-07817-3} {\bibfield
  {journal} {\bibinfo  {journal} {Nat. Commun.}\ }\textbf {\bibinfo {volume}
  {9}},\ \bibinfo {pages} {5384} (\bibinfo {year} {2018})},\ \Eprint
  {http://arxiv.org/abs/1611.01998} {1611.01998} \BibitemShut {NoStop}%
\bibitem [{\citenamefont {Avron}\ \emph {et~al.}(1988)\citenamefont {Avron},
  \citenamefont {Sadun}, \citenamefont {Segert},\ and\ \citenamefont
  {Simon}}]{Avron1988}%
  \BibitemOpen
  \bibfield  {author} {\bibinfo {author} {\bibfnamefont {J.~E.}\ \bibnamefont
  {Avron}}, \bibinfo {author} {\bibfnamefont {L.}~\bibnamefont {Sadun}},
  \bibinfo {author} {\bibfnamefont {J.}~\bibnamefont {Segert}}, \ and\ \bibinfo
  {author} {\bibfnamefont {B.}~\bibnamefont {Simon}},\ }\bibfield  {title}
  {\bibinfo {title} {\emph {{Topological Invariants in Fermi Systems with
  Time-Reversal Invariance}}},\ }\href {\doibase 10.1103/PhysRevLett.61.1329}
  {\bibfield  {journal} {\bibinfo  {journal} {Phys. Rev. Lett.}\ }\textbf
  {\bibinfo {volume} {61}},\ \bibinfo {pages} {1329} (\bibinfo {year}
  {1988})}\BibitemShut {NoStop}%
\bibitem [{\citenamefont {Fr{\"o}hlich}\ and\ \citenamefont
  {Perdini}(2000)}]{frohlich2000}%
  \BibitemOpen
  \bibfield  {author} {\bibinfo {author} {\bibfnamefont {J.}~\bibnamefont
  {Fr{\"o}hlich}}\ and\ \bibinfo {author} {\bibfnamefont {B.}~\bibnamefont
  {Perdini}},\ }\bibinfo {title} {\emph {New applications of the chiral
  anomaly}},\ in\ \href {\doibase 10.1142/9781848160224_0002} {\emph {\bibinfo
  {booktitle} {Mathematical Physics 2000}}}\ (\bibinfo  {publisher} {Imperial
  College Press, London, United Kingdom},\ \bibinfo {year} {2000})\ pp.\
  \bibinfo {pages} {9--47}\BibitemShut {NoStop}%
\bibitem [{\citenamefont {Zhang}\ and\ \citenamefont
  {Hu}(2001)}]{zhang2001four}%
  \BibitemOpen
  \bibfield  {author} {\bibinfo {author} {\bibfnamefont {S.-C.}\ \bibnamefont
  {Zhang}}\ and\ \bibinfo {author} {\bibfnamefont {J.}~\bibnamefont {Hu}},\
  }\bibfield  {title} {\bibinfo {title} {\emph {{A Four-Dimensional
  Generalization of the Quantum Hall Effect}}},\ }\href {\doibase
  10.1126/science.294.5543.823} {\bibfield  {journal} {\bibinfo  {journal}
  {Science}\ }\textbf {\bibinfo {volume} {294}},\ \bibinfo {pages} {823}
  (\bibinfo {year} {2001})}\BibitemShut {NoStop}%
\bibitem [{\citenamefont {Qi}\ \emph {et~al.}(2008)\citenamefont {Qi},
  \citenamefont {Hughes},\ and\ \citenamefont {Zhang}}]{qi2008}%
  \BibitemOpen
  \bibfield  {author} {\bibinfo {author} {\bibfnamefont {X.-L.}\ \bibnamefont
  {Qi}}, \bibinfo {author} {\bibfnamefont {T.~L.}\ \bibnamefont {Hughes}}, \
  and\ \bibinfo {author} {\bibfnamefont {S.-C.}\ \bibnamefont {Zhang}},\
  }\bibfield  {title} {\bibinfo {title} {\emph {Topological field theory of
  time-reversal invariant insulators}},\ }\href {\doibase
  10.1103/PhysRevB.78.195424} {\bibfield  {journal} {\bibinfo  {journal} {Phys.
  Rev. B}\ }\textbf {\bibinfo {volume} {78}},\ \bibinfo {pages} {195424}
  (\bibinfo {year} {2008})}\BibitemShut {NoStop}%
\bibitem [{\citenamefont {Edge}\ \emph {et~al.}(2012)\citenamefont {Edge},
  \citenamefont {Tworzyd{\l}o},\ and\ \citenamefont
  {Beenakker}}]{edge2012metallic}%
  \BibitemOpen
  \bibfield  {author} {\bibinfo {author} {\bibfnamefont {J.~M.}\ \bibnamefont
  {Edge}}, \bibinfo {author} {\bibfnamefont {J.}~\bibnamefont {Tworzyd{\l}o}},
  \ and\ \bibinfo {author} {\bibfnamefont {C.~W.~J.}\ \bibnamefont
  {Beenakker}},\ }\bibfield  {title} {\bibinfo {title} {\emph {{Metallic Phase
  of the Quantum Hall Effect in Four-Dimensional Space}}},\ }\href {\doibase
  10.1103/PhysRevLett.109.135701} {\bibfield  {journal} {\bibinfo  {journal}
  {Phys. Rev. Lett.}\ }\textbf {\bibinfo {volume} {109}},\ \bibinfo {pages}
  {135701} (\bibinfo {year} {2012})}\BibitemShut {NoStop}%
\bibitem [{\citenamefont {Kraus}\ \emph {et~al.}(2013)\citenamefont {Kraus},
  \citenamefont {Ringel},\ and\ \citenamefont {Zilberberg}}]{Kraus2013}%
  \BibitemOpen
  \bibfield  {author} {\bibinfo {author} {\bibfnamefont {Y.~E.}\ \bibnamefont
  {Kraus}}, \bibinfo {author} {\bibfnamefont {Z.}~\bibnamefont {Ringel}}, \
  and\ \bibinfo {author} {\bibfnamefont {O.}~\bibnamefont {Zilberberg}},\
  }\bibfield  {title} {\bibinfo {title} {\emph {Four-Dimensional Quantum Hall
  Effect in a Two-Dimensional Quasicrystal}},\ }\href {\doibase
  10.1103/PhysRevLett.111.226401} {\bibfield  {journal} {\bibinfo  {journal}
  {Phys. Rev. Lett.}\ }\textbf {\bibinfo {volume} {111}},\ \bibinfo {pages}
  {226401} (\bibinfo {year} {2013})}\BibitemShut {NoStop}%
\bibitem [{\citenamefont {Price}\ \emph {et~al.}(2015)\citenamefont {Price},
  \citenamefont {Zilberberg}, \citenamefont {Ozawa}, \citenamefont
  {Carusotto},\ and\ \citenamefont {Goldman}}]{Price2015}%
  \BibitemOpen
  \bibfield  {author} {\bibinfo {author} {\bibfnamefont {H.~M.}\ \bibnamefont
  {Price}}, \bibinfo {author} {\bibfnamefont {O.}~\bibnamefont {Zilberberg}},
  \bibinfo {author} {\bibfnamefont {T.}~\bibnamefont {Ozawa}}, \bibinfo
  {author} {\bibfnamefont {I.}~\bibnamefont {Carusotto}}, \ and\ \bibinfo
  {author} {\bibfnamefont {N.}~\bibnamefont {Goldman}},\ }\bibfield  {title}
  {\bibinfo {title} {\emph {Four-Dimensional Quantum Hall Effect with Ultracold
  Atoms}},\ }\href {\doibase 10.1103/PhysRevLett.115.195303} {\bibfield
  {journal} {\bibinfo  {journal} {Phys. Rev. Lett.}\ }\textbf {\bibinfo
  {volume} {115}},\ \bibinfo {pages} {195303} (\bibinfo {year}
  {2015})}\BibitemShut {NoStop}%
\bibitem [{\citenamefont {Price}\ \emph {et~al.}(2016)\citenamefont {Price},
  \citenamefont {Zilberberg}, \citenamefont {Ozawa}, \citenamefont
  {Carusotto},\ and\ \citenamefont {Goldman}}]{Price2016}%
  \BibitemOpen
  \bibfield  {author} {\bibinfo {author} {\bibfnamefont {H.~M.}\ \bibnamefont
  {Price}}, \bibinfo {author} {\bibfnamefont {O.}~\bibnamefont {Zilberberg}},
  \bibinfo {author} {\bibfnamefont {T.}~\bibnamefont {Ozawa}}, \bibinfo
  {author} {\bibfnamefont {I.}~\bibnamefont {Carusotto}}, \ and\ \bibinfo
  {author} {\bibfnamefont {N.}~\bibnamefont {Goldman}},\ }\bibfield  {title}
  {\bibinfo {title} {\emph {Measurement of Chern numbers through center-of-mass
  responses}},\ }\href {\doibase 10.1103/PhysRevB.93.245113} {\bibfield
  {journal} {\bibinfo  {journal} {Phys. Rev. B}\ }\textbf {\bibinfo {volume}
  {93}},\ \bibinfo {pages} {245113} (\bibinfo {year} {2016})}\BibitemShut
  {NoStop}%
\bibitem [{\citenamefont {Price}(2020)}]{Price2018}%
  \BibitemOpen
  \bibfield  {author} {\bibinfo {author} {\bibfnamefont {H.~M.}\ \bibnamefont
  {Price}},\ }\bibfield  {title} {\bibinfo {title} {\emph {Four-dimensional
  topological lattices through connectivity}},\ }\href {\doibase
  10.1103/PhysRevB.101.205141} {\bibfield  {journal} {\bibinfo  {journal}
  {Phys. Rev. B}\ }\textbf {\bibinfo {volume} {101}},\ \bibinfo {pages}
  {205141} (\bibinfo {year} {2020})}\BibitemShut {NoStop}%
\bibitem [{\citenamefont {{Lohse}}\ \emph {et~al.}(2018)\citenamefont
  {{Lohse}}, \citenamefont {{Schweizer}}, \citenamefont {{Price}},
  \citenamefont {{Zilberberg}},\ and\ \citenamefont {{Bloch}}}]{Lohse2018}%
  \BibitemOpen
  \bibfield  {author} {\bibinfo {author} {\bibfnamefont {M.}~\bibnamefont
  {{Lohse}}}, \bibinfo {author} {\bibfnamefont {C.}~\bibnamefont
  {{Schweizer}}}, \bibinfo {author} {\bibfnamefont {H.~M.}\ \bibnamefont
  {{Price}}}, \bibinfo {author} {\bibfnamefont {O.}~\bibnamefont
  {{Zilberberg}}}, \ and\ \bibinfo {author} {\bibfnamefont {I.}~\bibnamefont
  {{Bloch}}},\ }\bibfield  {title} {\bibinfo {title} {\emph {{Exploring 4D
  quantum Hall physics with a 2D topological charge pump}}},\ }\href {\doibase
  10.1038/nature25000} {\bibfield  {journal} {\bibinfo  {journal} {Nature}\
  }\textbf {\bibinfo {volume} {553}},\ \bibinfo {pages} {55} (\bibinfo {year}
  {2018})}\BibitemShut {NoStop}%
\bibitem [{\citenamefont {{Zilberberg}}\ \emph {et~al.}(2018)\citenamefont
  {{Zilberberg}}, \citenamefont {{Huang}}, \citenamefont {{Guglielmon}},
  \citenamefont {{Wang}}, \citenamefont {{Chen}}, \citenamefont {{Kraus}},\
  and\ \citenamefont {{Rechtsman}}}]{Zilberberg2018}%
  \BibitemOpen
  \bibfield  {author} {\bibinfo {author} {\bibfnamefont {O.}~\bibnamefont
  {{Zilberberg}}}, \bibinfo {author} {\bibfnamefont {S.}~\bibnamefont
  {{Huang}}}, \bibinfo {author} {\bibfnamefont {J.}~\bibnamefont
  {{Guglielmon}}}, \bibinfo {author} {\bibfnamefont {M.}~\bibnamefont
  {{Wang}}}, \bibinfo {author} {\bibfnamefont {K.}~\bibnamefont {{Chen}}},
  \bibinfo {author} {\bibfnamefont {Y.~E.}\ \bibnamefont {{Kraus}}}, \ and\
  \bibinfo {author} {\bibfnamefont {M.~C.}\ \bibnamefont {{Rechtsman}}},\
  }\bibfield  {title} {\bibinfo {title} {\emph {{Photonic topological boundary
  pumping as a probe of 4D quantum Hall physics}}},\ }\href {\doibase
  10.1038/nature25011} {\bibfield  {journal} {\bibinfo  {journal} {Nature}\
  }\textbf {\bibinfo {volume} {553}},\ \bibinfo {pages} {59} (\bibinfo {year}
  {2018})}\BibitemShut {NoStop}%
\bibitem [{\citenamefont {Sugawa}\ \emph {et~al.}(2018)\citenamefont {Sugawa},
  \citenamefont {Salces-Carcoba}, \citenamefont {Perry}, \citenamefont {Yue},\
  and\ \citenamefont {Spielman}}]{Sugawa2018}%
  \BibitemOpen
  \bibfield  {author} {\bibinfo {author} {\bibfnamefont {S.}~\bibnamefont
  {Sugawa}}, \bibinfo {author} {\bibfnamefont {F.}~\bibnamefont
  {Salces-Carcoba}}, \bibinfo {author} {\bibfnamefont {A.~R.}\ \bibnamefont
  {Perry}}, \bibinfo {author} {\bibfnamefont {Y.}~\bibnamefont {Yue}}, \ and\
  \bibinfo {author} {\bibfnamefont {I.~B.}\ \bibnamefont {Spielman}},\
  }\bibfield  {title} {\bibinfo {title} {\emph {Second Chern number of a
  quantum-simulated non-Abelian Yang monopole}},\ }\href {\doibase
  10.1126/science.aam9031} {\bibfield  {journal} {\bibinfo  {journal}
  {Science}\ }\textbf {\bibinfo {volume} {360}},\ \bibinfo {pages} {1429}
  (\bibinfo {year} {2018})}\BibitemShut {NoStop}%
\bibitem [{\citenamefont {Wang}\ \emph {et~al.}(2020)\citenamefont {Wang},
  \citenamefont {Price}, \citenamefont {Zhang},\ and\ \citenamefont
  {Chong}}]{wang2020circuit}%
  \BibitemOpen
  \bibfield  {author} {\bibinfo {author} {\bibfnamefont {Y.}~\bibnamefont
  {Wang}}, \bibinfo {author} {\bibfnamefont {H.~M.}\ \bibnamefont {Price}},
  \bibinfo {author} {\bibfnamefont {B.}~\bibnamefont {Zhang}}, \ and\ \bibinfo
  {author} {\bibfnamefont {Y.~D.}\ \bibnamefont {Chong}},\ }\bibfield  {title}
  {\bibinfo {title} {\emph {{Circuit implementation of a four-dimensional
  topological insulator}}},\ }\href {\doibase 10.1038/s41467-020-15940-3}
  {\bibfield  {journal} {\bibinfo  {journal} {Nat. Commun.}\ }\textbf {\bibinfo
  {volume} {11}},\ \bibinfo {pages} {2356} (\bibinfo {year}
  {2020})}\BibitemShut {NoStop}%
\bibitem [{\citenamefont {Chen}\ \emph {et~al.}(2021)\citenamefont {Chen},
  \citenamefont {Zhu}, \citenamefont {Tan}, \citenamefont {Wang},\ and\
  \citenamefont {Ma}}]{chen2021}%
  \BibitemOpen
  \bibfield  {author} {\bibinfo {author} {\bibfnamefont {Z.-G.}\ \bibnamefont
  {Chen}}, \bibinfo {author} {\bibfnamefont {W.}~\bibnamefont {Zhu}}, \bibinfo
  {author} {\bibfnamefont {Y.}~\bibnamefont {Tan}}, \bibinfo {author}
  {\bibfnamefont {L.}~\bibnamefont {Wang}}, \ and\ \bibinfo {author}
  {\bibfnamefont {G.}~\bibnamefont {Ma}},\ }\bibfield  {title} {\bibinfo
  {title} {\emph {Acoustic Realization of a Four-Dimensional Higher-Order Chern
  Insulator and Boundary-Modes Engineering}},\ }\href {\doibase
  10.1103/PhysRevX.11.011016} {\bibfield  {journal} {\bibinfo  {journal} {Phys.
  Rev. X}\ }\textbf {\bibinfo {volume} {11}},\ \bibinfo {pages} {011016}
  (\bibinfo {year} {2021})}\BibitemShut {NoStop}%
\bibitem [{\citenamefont {Petrides}\ and\ \citenamefont
  {Zilberberg}(2020)}]{Petrides2020}%
  \BibitemOpen
  \bibfield  {author} {\bibinfo {author} {\bibfnamefont {I.}~\bibnamefont
  {Petrides}}\ and\ \bibinfo {author} {\bibfnamefont {O.}~\bibnamefont
  {Zilberberg}},\ }\bibfield  {title} {\bibinfo {title} {\emph {{Higher-order
  topological insulators, topological pumps and the quantum Hall effect in high
  dimensions}}},\ }\href {\doibase 10.1103/PhysRevResearch.2.022049} {\bibfield
   {journal} {\bibinfo  {journal} {Phys. Rev. Res.}\ }\textbf {\bibinfo
  {volume} {2}},\ \bibinfo {pages} {022049} (\bibinfo {year}
  {2020})}\BibitemShut {NoStop}%
\bibitem [{\citenamefont {Goldman}\ \emph {et~al.}(2014)\citenamefont
  {Goldman}, \citenamefont {Juzeliunas}, \citenamefont {{\"{O}}hberg},\ and\
  \citenamefont {Spielman}}]{Goldman2014}%
  \BibitemOpen
  \bibfield  {author} {\bibinfo {author} {\bibfnamefont {N.}~\bibnamefont
  {Goldman}}, \bibinfo {author} {\bibfnamefont {G.}~\bibnamefont {Juzeliunas}},
  \bibinfo {author} {\bibfnamefont {P.}~\bibnamefont {{\"{O}}hberg}}, \ and\
  \bibinfo {author} {\bibfnamefont {I.~B.}\ \bibnamefont {Spielman}},\
  }\bibfield  {title} {\bibinfo {title} {\emph {Light-induced gauge fields for
  ultracold atoms}},\ }\href {\doibase 10.1088/0034-4885/77/12/126401}
  {\bibfield  {journal} {\bibinfo  {journal} {Rep. Prog. Phys.}\ }\textbf
  {\bibinfo {volume} {77}},\ \bibinfo {pages} {126401} (\bibinfo {year}
  {2014})}\BibitemShut {NoStop}%
\bibitem [{\citenamefont {Ozawa}\ \emph {et~al.}(2019)\citenamefont {Ozawa},
  \citenamefont {Price}, \citenamefont {Amo}, \citenamefont {Goldman},
  \citenamefont {Hafezi}, \citenamefont {Lu}, \citenamefont {Rechtsman},
  \citenamefont {Schuster}, \citenamefont {Simon}, \citenamefont {Zilberberg},\
  and\ \citenamefont {Carusotto}}]{Ozawa2018}%
  \BibitemOpen
  \bibfield  {author} {\bibinfo {author} {\bibfnamefont {T.}~\bibnamefont
  {Ozawa}}, \bibinfo {author} {\bibfnamefont {H.~M.}\ \bibnamefont {Price}},
  \bibinfo {author} {\bibfnamefont {A.}~\bibnamefont {Amo}}, \bibinfo {author}
  {\bibfnamefont {N.}~\bibnamefont {Goldman}}, \bibinfo {author} {\bibfnamefont
  {M.}~\bibnamefont {Hafezi}}, \bibinfo {author} {\bibfnamefont
  {L.}~\bibnamefont {Lu}}, \bibinfo {author} {\bibfnamefont {M.~C.}\
  \bibnamefont {Rechtsman}}, \bibinfo {author} {\bibfnamefont {D.}~\bibnamefont
  {Schuster}}, \bibinfo {author} {\bibfnamefont {J.}~\bibnamefont {Simon}},
  \bibinfo {author} {\bibfnamefont {O.}~\bibnamefont {Zilberberg}}, \ and\
  \bibinfo {author} {\bibfnamefont {I.}~\bibnamefont {Carusotto}},\ }\bibfield
  {title} {\bibinfo {title} {\emph {Topological photonics}},\ }\href {\doibase
  10.1103/RevModPhys.91.015006} {\bibfield  {journal} {\bibinfo  {journal}
  {Rev. Mod. Phys.}\ }\textbf {\bibinfo {volume} {91}},\ \bibinfo {pages}
  {015006} (\bibinfo {year} {2019})}\BibitemShut {NoStop}%
\bibitem [{\citenamefont {Cooper}\ \emph {et~al.}(2019)\citenamefont {Cooper},
  \citenamefont {Dalibard},\ and\ \citenamefont {Spielman}}]{Cooper2018}%
  \BibitemOpen
  \bibfield  {author} {\bibinfo {author} {\bibfnamefont {N.~R.}\ \bibnamefont
  {Cooper}}, \bibinfo {author} {\bibfnamefont {J.}~\bibnamefont {Dalibard}}, \
  and\ \bibinfo {author} {\bibfnamefont {I.~B.}\ \bibnamefont {Spielman}},\
  }\bibfield  {title} {\bibinfo {title} {\emph {Topological bands for ultracold
  atoms}},\ }\href {\doibase 10.1103/RevModPhys.91.015005} {\bibfield
  {journal} {\bibinfo  {journal} {Rev. Mod. Phys.}\ }\textbf {\bibinfo {volume}
  {91}},\ \bibinfo {pages} {015005} (\bibinfo {year} {2019})}\BibitemShut
  {NoStop}%
\bibitem [{\citenamefont {Aidelsburger}\ \emph {et~al.}(2014)\citenamefont
  {Aidelsburger}, \citenamefont {Lohse}, \citenamefont {Schweizer},
  \citenamefont {Atala}, \citenamefont {Barreiro}, \citenamefont
  {Nascimb{\`{e}}ne}, \citenamefont {Cooper}, \citenamefont {Bloch},\ and\
  \citenamefont {Goldman}}]{Aidelsburger2014}%
  \BibitemOpen
  \bibfield  {author} {\bibinfo {author} {\bibfnamefont {M.}~\bibnamefont
  {Aidelsburger}}, \bibinfo {author} {\bibfnamefont {M.}~\bibnamefont {Lohse}},
  \bibinfo {author} {\bibfnamefont {C.}~\bibnamefont {Schweizer}}, \bibinfo
  {author} {\bibfnamefont {M.}~\bibnamefont {Atala}}, \bibinfo {author}
  {\bibfnamefont {J.~T.}\ \bibnamefont {Barreiro}}, \bibinfo {author}
  {\bibfnamefont {S.}~\bibnamefont {Nascimb{\`{e}}ne}}, \bibinfo {author}
  {\bibfnamefont {N.~R.}\ \bibnamefont {Cooper}}, \bibinfo {author}
  {\bibfnamefont {I.}~\bibnamefont {Bloch}}, \ and\ \bibinfo {author}
  {\bibfnamefont {N.}~\bibnamefont {Goldman}},\ }\bibfield  {title} {\bibinfo
  {title} {\emph {Measuring the Chern number of Hofstadter bands with ultracold
  bosonic atoms}},\ }\href {\doibase 10.1038/nphys3171} {\bibfield  {journal}
  {\bibinfo  {journal} {Nat. Phys.}\ }\textbf {\bibinfo {volume} {11}},\
  \bibinfo {pages} {162} (\bibinfo {year} {2014})}\BibitemShut {NoStop}%
\bibitem [{\citenamefont {Asteria}\ \emph {et~al.}(2019)\citenamefont
  {Asteria}, \citenamefont {Tran}, \citenamefont {Ozawa}, \citenamefont
  {Tarnowski}, \citenamefont {Rem}, \citenamefont {Fl{\"a}schner},
  \citenamefont {Sengstock}, \citenamefont {Goldman},\ and\ \citenamefont
  {Weitenberg}}]{Asteria19}%
  \BibitemOpen
  \bibfield  {author} {\bibinfo {author} {\bibfnamefont {L.}~\bibnamefont
  {Asteria}}, \bibinfo {author} {\bibfnamefont {D.~T.}\ \bibnamefont {Tran}},
  \bibinfo {author} {\bibfnamefont {T.}~\bibnamefont {Ozawa}}, \bibinfo
  {author} {\bibfnamefont {M.}~\bibnamefont {Tarnowski}}, \bibinfo {author}
  {\bibfnamefont {B.~S.}\ \bibnamefont {Rem}}, \bibinfo {author} {\bibfnamefont
  {N.}~\bibnamefont {Fl{\"a}schner}}, \bibinfo {author} {\bibfnamefont
  {K.}~\bibnamefont {Sengstock}}, \bibinfo {author} {\bibfnamefont
  {N.}~\bibnamefont {Goldman}}, \ and\ \bibinfo {author} {\bibfnamefont
  {C.}~\bibnamefont {Weitenberg}},\ }\bibfield  {title} {\bibinfo {title}
  {\emph {Measuring quantized circular dichroism in ultracold topological
  matter}},\ }\href {\doibase 10.1038/s41567-019-0417-8} {\bibfield  {journal}
  {\bibinfo  {journal} {Nat. Phys.}\ }\textbf {\bibinfo {volume} {15}},\
  \bibinfo {pages} {449} (\bibinfo {year} {2019})}\BibitemShut {NoStop}%
\bibitem [{\citenamefont {Kraus}\ \emph {et~al.}(2012)\citenamefont {Kraus},
  \citenamefont {Lahini}, \citenamefont {Ringel}, \citenamefont {Verbin},\ and\
  \citenamefont {Zilberberg}}]{Kraus2012}%
  \BibitemOpen
  \bibfield  {author} {\bibinfo {author} {\bibfnamefont {Y.~E.}\ \bibnamefont
  {Kraus}}, \bibinfo {author} {\bibfnamefont {Y.}~\bibnamefont {Lahini}},
  \bibinfo {author} {\bibfnamefont {Z.}~\bibnamefont {Ringel}}, \bibinfo
  {author} {\bibfnamefont {M.}~\bibnamefont {Verbin}}, \ and\ \bibinfo {author}
  {\bibfnamefont {O.}~\bibnamefont {Zilberberg}},\ }\bibfield  {title}
  {\bibinfo {title} {\emph {Topological States and Adiabatic Pumping in
  Quasicrystals}},\ }\href {\doibase 10.1103/PhysRevLett.109.106402} {\bibfield
   {journal} {\bibinfo  {journal} {Phys. Rev. Lett.}\ }\textbf {\bibinfo
  {volume} {109}},\ \bibinfo {pages} {106402} (\bibinfo {year}
  {2012})}\BibitemShut {NoStop}%
\bibitem [{\citenamefont {Mittal}\ \emph {et~al.}(2019)\citenamefont {Mittal},
  \citenamefont {Orre}, \citenamefont {Leykam}, \citenamefont {Chong},\ and\
  \citenamefont {Hafezi}}]{Mittal2019}%
  \BibitemOpen
  \bibfield  {author} {\bibinfo {author} {\bibfnamefont {S.}~\bibnamefont
  {Mittal}}, \bibinfo {author} {\bibfnamefont {V.~V.}\ \bibnamefont {Orre}},
  \bibinfo {author} {\bibfnamefont {D.}~\bibnamefont {Leykam}}, \bibinfo
  {author} {\bibfnamefont {Y.~D.}\ \bibnamefont {Chong}}, \ and\ \bibinfo
  {author} {\bibfnamefont {M.}~\bibnamefont {Hafezi}},\ }\bibfield  {title}
  {\bibinfo {title} {\emph {{Photonic Anomalous Quantum Hall Effect}}},\ }\href
  {\doibase 10.1103/PhysRevLett.123.043201} {\bibfield  {journal} {\bibinfo
  {journal} {Phys. Rev. Lett.}\ }\textbf {\bibinfo {volume} {123}},\ \bibinfo
  {pages} {043201} (\bibinfo {year} {2019})}\BibitemShut {NoStop}%
\bibitem [{\citenamefont {D'Errico}\ \emph {et~al.}(2020)\citenamefont
  {D'Errico}, \citenamefont {Cardano}, \citenamefont {Maffei}, \citenamefont
  {Dauphin}, \citenamefont {Barboza}, \citenamefont {Esposito}, \citenamefont
  {Piccirillo}, \citenamefont {Lewenstein}, \citenamefont {Massignan},\ and\
  \citenamefont {Marrucci}}]{DErrico20}%
  \BibitemOpen
  \bibfield  {author} {\bibinfo {author} {\bibfnamefont {A.}~\bibnamefont
  {D'Errico}}, \bibinfo {author} {\bibfnamefont {F.}~\bibnamefont {Cardano}},
  \bibinfo {author} {\bibfnamefont {M.}~\bibnamefont {Maffei}}, \bibinfo
  {author} {\bibfnamefont {A.}~\bibnamefont {Dauphin}}, \bibinfo {author}
  {\bibfnamefont {R.}~\bibnamefont {Barboza}}, \bibinfo {author} {\bibfnamefont
  {C.}~\bibnamefont {Esposito}}, \bibinfo {author} {\bibfnamefont
  {B.}~\bibnamefont {Piccirillo}}, \bibinfo {author} {\bibfnamefont
  {M.}~\bibnamefont {Lewenstein}}, \bibinfo {author} {\bibfnamefont
  {P.}~\bibnamefont {Massignan}}, \ and\ \bibinfo {author} {\bibfnamefont
  {L.}~\bibnamefont {Marrucci}},\ }\bibfield  {title} {\bibinfo {title} {\emph
  {Two-dimensional topological quantum walks in the momentum space of
  structured light}},\ }\href {\doibase 10.1364/OPTICA.365028} {\bibfield
  {journal} {\bibinfo  {journal} {Optica}\ }\textbf {\bibinfo {volume} {7}},\
  \bibinfo {pages} {108} (\bibinfo {year} {2020})}\BibitemShut {NoStop}%
\bibitem [{\citenamefont {Ni}\ \emph {et~al.}(2019)\citenamefont {Ni},
  \citenamefont {Chen}, \citenamefont {Weiner}, \citenamefont {Apigo},
  \citenamefont {Prodan}, \citenamefont {Al{\`u}}, \citenamefont {Prodan},\
  and\ \citenamefont {Khanikaev}}]{Ni2019}%
  \BibitemOpen
  \bibfield  {author} {\bibinfo {author} {\bibfnamefont {X.}~\bibnamefont
  {Ni}}, \bibinfo {author} {\bibfnamefont {K.}~\bibnamefont {Chen}}, \bibinfo
  {author} {\bibfnamefont {M.}~\bibnamefont {Weiner}}, \bibinfo {author}
  {\bibfnamefont {D.~J.}\ \bibnamefont {Apigo}}, \bibinfo {author}
  {\bibfnamefont {C.}~\bibnamefont {Prodan}}, \bibinfo {author} {\bibfnamefont
  {A.}~\bibnamefont {Al{\`u}}}, \bibinfo {author} {\bibfnamefont
  {E.}~\bibnamefont {Prodan}}, \ and\ \bibinfo {author} {\bibfnamefont {A.~B.}\
  \bibnamefont {Khanikaev}},\ }\bibfield  {title} {\bibinfo {title} {\emph
  {Observation of Hofstadter butterfly and topological edge states in
  reconfigurable quasi-periodic acoustic crystals}},\ }\href {\doibase
  10.1038/s42005-019-0151-7} {\bibfield  {journal} {\bibinfo  {journal}
  {Commun. Phys.}\ }\textbf {\bibinfo {volume} {2}},\ \bibinfo {pages} {55}
  (\bibinfo {year} {2019})}\BibitemShut {NoStop}%
\bibitem [{\citenamefont {Tang}\ \emph {et~al.}(2019)\citenamefont {Tang},
  \citenamefont {Ren}, \citenamefont {Wang}, \citenamefont {Zhong},
  \citenamefont {Schneeloch}, \citenamefont {Yang}, \citenamefont {Yang},
  \citenamefont {Lee}, \citenamefont {Gu}, \citenamefont {Qiao},\ and\
  \citenamefont {Zhang}}]{Tang2019}%
  \BibitemOpen
  \bibfield  {author} {\bibinfo {author} {\bibfnamefont {F.}~\bibnamefont
  {Tang}}, \bibinfo {author} {\bibfnamefont {Y.}~\bibnamefont {Ren}}, \bibinfo
  {author} {\bibfnamefont {P.}~\bibnamefont {Wang}}, \bibinfo {author}
  {\bibfnamefont {R.}~\bibnamefont {Zhong}}, \bibinfo {author} {\bibfnamefont
  {J.}~\bibnamefont {Schneeloch}}, \bibinfo {author} {\bibfnamefont {S.~A.}\
  \bibnamefont {Yang}}, \bibinfo {author} {\bibfnamefont {K.}~\bibnamefont
  {Yang}}, \bibinfo {author} {\bibfnamefont {P.~A.}\ \bibnamefont {Lee}},
  \bibinfo {author} {\bibfnamefont {G.}~\bibnamefont {Gu}}, \bibinfo {author}
  {\bibfnamefont {Z.}~\bibnamefont {Qiao}}, \ and\ \bibinfo {author}
  {\bibfnamefont {L.}~\bibnamefont {Zhang}},\ }\bibfield  {title} {\bibinfo
  {title} {\emph {Three-dimensional quantum Hall effect and metal-insulator
  transition in {ZrTe}$_5$}},\ }\href {\doibase 10.1038/s41586-019-1180-9}
  {\bibfield  {journal} {\bibinfo  {journal} {Nature}\ }\textbf {\bibinfo
  {volume} {569}},\ \bibinfo {pages} {537} (\bibinfo {year}
  {2019})}\BibitemShut {NoStop}%
\bibitem [{\citenamefont {Krinner}\ \emph {et~al.}(2014)\citenamefont
  {Krinner}, \citenamefont {Stadler}, \citenamefont {Husmann}, \citenamefont
  {Brantut},\ and\ \citenamefont {Esslinger}}]{Krinner2014}%
  \BibitemOpen
  \bibfield  {author} {\bibinfo {author} {\bibfnamefont {S.}~\bibnamefont
  {Krinner}}, \bibinfo {author} {\bibfnamefont {D.}~\bibnamefont {Stadler}},
  \bibinfo {author} {\bibfnamefont {D.}~\bibnamefont {Husmann}}, \bibinfo
  {author} {\bibfnamefont {J.-P.}\ \bibnamefont {Brantut}}, \ and\ \bibinfo
  {author} {\bibfnamefont {T.}~\bibnamefont {Esslinger}},\ }\bibfield  {title}
  {\bibinfo {title} {\emph {Observation of quantized conductance in neutral
  matter}},\ }\href {\doibase 10.1038/nature14049} {\bibfield  {journal}
  {\bibinfo  {journal} {Nature}\ }\textbf {\bibinfo {volume} {517}},\ \bibinfo
  {pages} {64} (\bibinfo {year} {2014})}\BibitemShut {NoStop}%
\bibitem [{\citenamefont {Umucal\ifmmode \imath \else~\i \fi{}lar}\ \emph
  {et~al.}(2008)\citenamefont {Umucal\ifmmode \imath \else~\i \fi{}lar},
  \citenamefont {Zhai},\ and\ \citenamefont {Oktel}}]{oktel2008}%
  \BibitemOpen
  \bibfield  {author} {\bibinfo {author} {\bibfnamefont {R.~O.}\ \bibnamefont
  {Umucal\ifmmode \imath \else~\i \fi{}lar}}, \bibinfo {author} {\bibfnamefont
  {H.}~\bibnamefont {Zhai}}, \ and\ \bibinfo {author} {\bibfnamefont {M.~O.}\
  \bibnamefont {Oktel}},\ }\bibfield  {title} {\bibinfo {title} {\emph {Trapped
  Fermi Gases in Rotating Optical Lattices: Realization and Detection of the
  Topological Hofstadter Insulator}},\ }\href {\doibase
  10.1103/PhysRevLett.100.070402} {\bibfield  {journal} {\bibinfo  {journal}
  {Phys. Rev. Lett.}\ }\textbf {\bibinfo {volume} {100}},\ \bibinfo {pages}
  {070402} (\bibinfo {year} {2008})}\BibitemShut {NoStop}%
\bibitem [{\citenamefont {Bardyn}\ \emph {et~al.}(2014)\citenamefont {Bardyn},
  \citenamefont {Huber},\ and\ \citenamefont {Zilberberg}}]{Bardyn:2014}%
  \BibitemOpen
  \bibfield  {author} {\bibinfo {author} {\bibfnamefont {C.-E.}\ \bibnamefont
  {Bardyn}}, \bibinfo {author} {\bibfnamefont {S.~D.}\ \bibnamefont {Huber}}, \
  and\ \bibinfo {author} {\bibfnamefont {O.}~\bibnamefont {Zilberberg}},\
  }\bibfield  {title} {\bibinfo {title} {\emph {Measuring topological
  invariants in small photonic lattices}},\ }\href
  {https://iopscience.iop.org/article/10.1088/1367-2630/16/12/123013}
  {\bibfield  {journal} {\bibinfo  {journal} {New J. Phys.}\ }\textbf {\bibinfo
  {volume} {16}},\ \bibinfo {pages} {123013} (\bibinfo {year}
  {2014})}\BibitemShut {NoStop}%
\bibitem [{\citenamefont {Tran}\ \emph {et~al.}(2017)\citenamefont {Tran},
  \citenamefont {Dauphin}, \citenamefont {Grushin}, \citenamefont {Zoller},\
  and\ \citenamefont {Goldman}}]{Tran17}%
  \BibitemOpen
  \bibfield  {author} {\bibinfo {author} {\bibfnamefont {D.~T.}\ \bibnamefont
  {Tran}}, \bibinfo {author} {\bibfnamefont {A.}~\bibnamefont {Dauphin}},
  \bibinfo {author} {\bibfnamefont {A.~G.}\ \bibnamefont {Grushin}}, \bibinfo
  {author} {\bibfnamefont {P.}~\bibnamefont {Zoller}}, \ and\ \bibinfo {author}
  {\bibfnamefont {N.}~\bibnamefont {Goldman}},\ }\bibfield  {title} {\bibinfo
  {title} {\emph {{Probing topology by
  {\textquotedblleft}heating{\textquotedblright}: Quantized circular dichroism
  in ultracold atoms}}},\ }\href {\doibase 10.1126/sciadv.1701207} {\bibfield
  {journal} {\bibinfo  {journal} {Sci. Adv.}\ }\textbf {\bibinfo {volume}
  {3}},\ \bibinfo {pages} {e1701207} (\bibinfo {year} {2017})}\BibitemShut
  {NoStop}%
\bibitem [{\citenamefont {Wang}\ \emph {et~al.}(2017)\citenamefont {Wang},
  \citenamefont {Zhang}, \citenamefont {Chen}, \citenamefont {Yu},\ and\
  \citenamefont {Zhai}}]{wang17}%
  \BibitemOpen
  \bibfield  {author} {\bibinfo {author} {\bibfnamefont {C.}~\bibnamefont
  {Wang}}, \bibinfo {author} {\bibfnamefont {P.}~\bibnamefont {Zhang}},
  \bibinfo {author} {\bibfnamefont {X.}~\bibnamefont {Chen}}, \bibinfo {author}
  {\bibfnamefont {J.}~\bibnamefont {Yu}}, \ and\ \bibinfo {author}
  {\bibfnamefont {H.}~\bibnamefont {Zhai}},\ }\bibfield  {title} {\bibinfo
  {title} {\emph {Scheme to Measure the Topological Number of a Chern Insulator
  from Quench Dynamics}},\ }\href {\doibase 10.1103/PhysRevLett.118.185701}
  {\bibfield  {journal} {\bibinfo  {journal} {Phys. Rev. Lett.}\ }\textbf
  {\bibinfo {volume} {118}},\ \bibinfo {pages} {185701} (\bibinfo {year}
  {2017})}\BibitemShut {NoStop}%
\bibitem [{\citenamefont {Cardano}\ \emph {et~al.}(2017)\citenamefont
  {Cardano}, \citenamefont {D{'}Errico}, \citenamefont {Dauphin}, \citenamefont
  {Maffei}, \citenamefont {Piccirillo}, \citenamefont {de~Lisio}, \citenamefont
  {De~Filippis}, \citenamefont {Cataudella}, \citenamefont {Santamato},
  \citenamefont {Marrucci}, \citenamefont {Lewenstein},\ and\ \citenamefont
  {Massignan}}]{Cardano2017}%
  \BibitemOpen
  \bibfield  {author} {\bibinfo {author} {\bibfnamefont {F.}~\bibnamefont
  {Cardano}}, \bibinfo {author} {\bibfnamefont {A.}~\bibnamefont {D{'}Errico}},
  \bibinfo {author} {\bibfnamefont {A.}~\bibnamefont {Dauphin}}, \bibinfo
  {author} {\bibfnamefont {M.}~\bibnamefont {Maffei}}, \bibinfo {author}
  {\bibfnamefont {B.}~\bibnamefont {Piccirillo}}, \bibinfo {author}
  {\bibfnamefont {C.}~\bibnamefont {de~Lisio}}, \bibinfo {author}
  {\bibfnamefont {G.}~\bibnamefont {De~Filippis}}, \bibinfo {author}
  {\bibfnamefont {V.}~\bibnamefont {Cataudella}}, \bibinfo {author}
  {\bibfnamefont {E.}~\bibnamefont {Santamato}}, \bibinfo {author}
  {\bibfnamefont {L.}~\bibnamefont {Marrucci}}, \bibinfo {author}
  {\bibfnamefont {M.}~\bibnamefont {Lewenstein}}, \ and\ \bibinfo {author}
  {\bibfnamefont {P.}~\bibnamefont {Massignan}},\ }\bibfield  {title} {\bibinfo
  {title} {\emph {{Detection of Zak phases and topological invariants in a
  chiral quantum walk of twisted photons}}},\ }\href {\doibase
  10.1038/ncomms15516} {\bibfield  {journal} {\bibinfo  {journal} {Nat.
  Commun.}\ }\textbf {\bibinfo {volume} {8}},\ \bibinfo {pages} {1} (\bibinfo
  {year} {2017})}\BibitemShut {NoStop}%
\bibitem [{\citenamefont {Tarnowski}\ \emph {et~al.}(2019)\citenamefont
  {Tarnowski}, \citenamefont {{\"U}nal}, \citenamefont {Fl{\"a}schner},
  \citenamefont {Rem}, \citenamefont {Eckardt}, \citenamefont {Sengstock},\
  and\ \citenamefont {Weitenberg}}]{Tarnowski19}%
  \BibitemOpen
  \bibfield  {author} {\bibinfo {author} {\bibfnamefont {M.}~\bibnamefont
  {Tarnowski}}, \bibinfo {author} {\bibfnamefont {F.~N.}\ \bibnamefont
  {{\"U}nal}}, \bibinfo {author} {\bibfnamefont {N.}~\bibnamefont
  {Fl{\"a}schner}}, \bibinfo {author} {\bibfnamefont {B.~S.}\ \bibnamefont
  {Rem}}, \bibinfo {author} {\bibfnamefont {A.}~\bibnamefont {Eckardt}},
  \bibinfo {author} {\bibfnamefont {K.}~\bibnamefont {Sengstock}}, \ and\
  \bibinfo {author} {\bibfnamefont {C.}~\bibnamefont {Weitenberg}},\ }\bibfield
   {title} {\bibinfo {title} {\emph {Measuring topology from dynamics by
  obtaining the Chern number from a linking number}},\ }\href {\doibase
  10.1038/s41467-019-09668-y} {\bibfield  {journal} {\bibinfo  {journal} {Nat.
  Commun.}\ }\textbf {\bibinfo {volume} {10}},\ \bibinfo {pages} {1728}
  (\bibinfo {year} {2019})}\BibitemShut {NoStop}%
\bibitem [{\citenamefont {Wannier}(1978)}]{Wannier1978}%
  \BibitemOpen
  \bibfield  {author} {\bibinfo {author} {\bibfnamefont {G.~H.}\ \bibnamefont
  {Wannier}},\ }\bibfield  {title} {\bibinfo {title} {\emph {A Result Not
  Dependent on Rationality for Bloch Electrons in a Magnetic Field}},\ }\href
  {\doibase 10.1002/pssb.2220880243} {\bibfield  {journal} {\bibinfo  {journal}
  {Phys. Status Solidi (B)}\ }\textbf {\bibinfo {volume} {88}},\ \bibinfo
  {pages} {757} (\bibinfo {year} {1978})}\BibitemShut {NoStop}%
\bibitem [{\citenamefont {Price}\ and\ \citenamefont
  {Cooper}(2012)}]{Price2012}%
  \BibitemOpen
  \bibfield  {author} {\bibinfo {author} {\bibfnamefont {H.~M.}\ \bibnamefont
  {Price}}\ and\ \bibinfo {author} {\bibfnamefont {N.~R.}\ \bibnamefont
  {Cooper}},\ }\bibfield  {title} {\bibinfo {title} {\emph {Mapping the Berry
  curvature from semiclassical dynamics in optical lattices}},\ }\href
  {\doibase 10.1103/PhysRevA.85.033620} {\bibfield  {journal} {\bibinfo
  {journal} {Phys. Rev. A}\ }\textbf {\bibinfo {volume} {85}},\ \bibinfo
  {pages} {033620} (\bibinfo {year} {2012})}\BibitemShut {NoStop}%
\bibitem [{\citenamefont {Dauphin}\ and\ \citenamefont
  {Goldman}(2013)}]{Dauphin2013}%
  \BibitemOpen
  \bibfield  {author} {\bibinfo {author} {\bibfnamefont {A.}~\bibnamefont
  {Dauphin}}\ and\ \bibinfo {author} {\bibfnamefont {N.}~\bibnamefont
  {Goldman}},\ }\bibfield  {title} {\bibinfo {title} {\emph {Extracting the
  Chern Number from the Dynamics of a Fermi Gas: Implementing a Quantum Hall
  Bar for Cold Atoms}},\ }\href {\doibase 10.1103/PhysRevLett.111.135302}
  {\bibfield  {journal} {\bibinfo  {journal} {Phys. Rev. Lett.}\ }\textbf
  {\bibinfo {volume} {111}},\ \bibinfo {pages} {135302} (\bibinfo {year}
  {2013})}\BibitemShut {NoStop}%
\bibitem [{\citenamefont {Thouless}(1983)}]{Thouless1983}%
  \BibitemOpen
  \bibfield  {author} {\bibinfo {author} {\bibfnamefont {D.~J.}\ \bibnamefont
  {Thouless}},\ }\bibfield  {title} {\bibinfo {title} {\emph {Quantization of
  particle transport}},\ }\href {\doibase 10.1103/PhysRevB.27.6083} {\bibfield
  {journal} {\bibinfo  {journal} {Phys. Rev. B}\ }\textbf {\bibinfo {volume}
  {27}},\ \bibinfo {pages} {6083} (\bibinfo {year} {1983})}\BibitemShut
  {NoStop}%
\bibitem [{\citenamefont {Lohse}\ \emph {et~al.}(2015)\citenamefont {Lohse},
  \citenamefont {Schweizer}, \citenamefont {Zilberberg}, \citenamefont
  {Aidelsburger},\ and\ \citenamefont {Bloch}}]{Lohse2015}%
  \BibitemOpen
  \bibfield  {author} {\bibinfo {author} {\bibfnamefont {M.}~\bibnamefont
  {Lohse}}, \bibinfo {author} {\bibfnamefont {C.}~\bibnamefont {Schweizer}},
  \bibinfo {author} {\bibfnamefont {O.}~\bibnamefont {Zilberberg}}, \bibinfo
  {author} {\bibfnamefont {M.}~\bibnamefont {Aidelsburger}}, \ and\ \bibinfo
  {author} {\bibfnamefont {I.}~\bibnamefont {Bloch}},\ }\bibfield  {title}
  {\bibinfo {title} {\emph {A Thouless quantum pump with ultracold bosonic
  atoms in an optical superlattice}},\ }\href {\doibase 10.1038/nphys3584}
  {\bibfield  {journal} {\bibinfo  {journal} {Nat. Phys.}\ }\textbf {\bibinfo
  {volume} {12}},\ \bibinfo {pages} {350} (\bibinfo {year} {2015})}\BibitemShut
  {NoStop}%
\bibitem [{\citenamefont {Nakajima}\ \emph {et~al.}(2016)\citenamefont
  {Nakajima}, \citenamefont {Tomita}, \citenamefont {Taie}, \citenamefont
  {Ichinose}, \citenamefont {Ozawa}, \citenamefont {Wang}, \citenamefont
  {Troyer},\ and\ \citenamefont {Takahashi}}]{Nakajima2016}%
  \BibitemOpen
  \bibfield  {author} {\bibinfo {author} {\bibfnamefont {S.}~\bibnamefont
  {Nakajima}}, \bibinfo {author} {\bibfnamefont {T.}~\bibnamefont {Tomita}},
  \bibinfo {author} {\bibfnamefont {S.}~\bibnamefont {Taie}}, \bibinfo {author}
  {\bibfnamefont {T.}~\bibnamefont {Ichinose}}, \bibinfo {author}
  {\bibfnamefont {H.}~\bibnamefont {Ozawa}}, \bibinfo {author} {\bibfnamefont
  {L.}~\bibnamefont {Wang}}, \bibinfo {author} {\bibfnamefont {M.}~\bibnamefont
  {Troyer}}, \ and\ \bibinfo {author} {\bibfnamefont {Y.}~\bibnamefont
  {Takahashi}},\ }\bibfield  {title} {\bibinfo {title} {\emph {Topological
  Thouless pumping of ultracold~fermions}},\ }\href {\doibase
  10.1038/nphys3622} {\bibfield  {journal} {\bibinfo  {journal} {Nat. Phys.}\
  }\textbf {\bibinfo {volume} {12}},\ \bibinfo {pages} {296} (\bibinfo {year}
  {2016})}\BibitemShut {NoStop}%
\bibitem [{\citenamefont {Shechtman}\ \emph {et~al.}(1984)\citenamefont
  {Shechtman}, \citenamefont {Blech}, \citenamefont {Gratias},\ and\
  \citenamefont {Cahn}}]{Shechtman1984}%
  \BibitemOpen
  \bibfield  {author} {\bibinfo {author} {\bibfnamefont {D.}~\bibnamefont
  {Shechtman}}, \bibinfo {author} {\bibfnamefont {I.}~\bibnamefont {Blech}},
  \bibinfo {author} {\bibfnamefont {D.}~\bibnamefont {Gratias}}, \ and\
  \bibinfo {author} {\bibfnamefont {J.~W.}\ \bibnamefont {Cahn}},\ }\bibfield
  {title} {\bibinfo {title} {\emph {Metallic Phase with Long-Range
  Orientational Order and No Translational Symmetry}},\ }\href {\doibase
  10.1103/PhysRevLett.53.1951} {\bibfield  {journal} {\bibinfo  {journal}
  {Phys. Rev. Lett.}\ }\textbf {\bibinfo {volume} {53}},\ \bibinfo {pages}
  {1951} (\bibinfo {year} {1984})}\BibitemShut {NoStop}%
\bibitem [{\citenamefont {Prodan}(2015)}]{Prodan2015}%
  \BibitemOpen
  \bibfield  {author} {\bibinfo {author} {\bibfnamefont {E.}~\bibnamefont
  {Prodan}},\ }\bibfield  {title} {\bibinfo {title} {\emph {Virtual topological
  insulators with real quantized physics}},\ }\href {\doibase
  10.1103/PhysRevB.91.245104} {\bibfield  {journal} {\bibinfo  {journal} {Phys.
  Rev. B}\ }\textbf {\bibinfo {volume} {91}},\ \bibinfo {pages} {245104}
  (\bibinfo {year} {2015})}\BibitemShut {NoStop}%
\bibitem [{\citenamefont {Dareau}\ \emph {et~al.}(2017)\citenamefont {Dareau},
  \citenamefont {Levy}, \citenamefont {Aguilera}, \citenamefont {Bouganne},
  \citenamefont {Akkermans}, \citenamefont {Gerbier},\ and\ \citenamefont
  {Beugnon}}]{Dareau2017}%
  \BibitemOpen
  \bibfield  {author} {\bibinfo {author} {\bibfnamefont {A.}~\bibnamefont
  {Dareau}}, \bibinfo {author} {\bibfnamefont {E.}~\bibnamefont {Levy}},
  \bibinfo {author} {\bibfnamefont {M.~B.}\ \bibnamefont {Aguilera}}, \bibinfo
  {author} {\bibfnamefont {R.}~\bibnamefont {Bouganne}}, \bibinfo {author}
  {\bibfnamefont {E.}~\bibnamefont {Akkermans}}, \bibinfo {author}
  {\bibfnamefont {F.}~\bibnamefont {Gerbier}}, \ and\ \bibinfo {author}
  {\bibfnamefont {J.}~\bibnamefont {Beugnon}},\ }\bibfield  {title} {\bibinfo
  {title} {\emph {Revealing the Topology of Quasicrystals with a Diffraction
  Experiment}},\ }\href {\doibase 10.1103/PhysRevLett.119.215304} {\bibfield
  {journal} {\bibinfo  {journal} {Phys. Rev. Lett.}\ }\textbf {\bibinfo
  {volume} {119}},\ \bibinfo {pages} {215304} (\bibinfo {year}
  {2017})}\BibitemShut {NoStop}%
\bibitem [{\citenamefont {Tanese}\ \emph {et~al.}(2014)\citenamefont {Tanese},
  \citenamefont {Gurevich}, \citenamefont {Baboux}, \citenamefont {Jacqmin},
  \citenamefont {Lema\^{\i}tre}, \citenamefont {Galopin}, \citenamefont
  {Sagnes}, \citenamefont {Amo}, \citenamefont {Bloch},\ and\ \citenamefont
  {Akkermans}}]{Tanese2014}%
  \BibitemOpen
  \bibfield  {author} {\bibinfo {author} {\bibfnamefont {D.}~\bibnamefont
  {Tanese}}, \bibinfo {author} {\bibfnamefont {E.}~\bibnamefont {Gurevich}},
  \bibinfo {author} {\bibfnamefont {F.}~\bibnamefont {Baboux}}, \bibinfo
  {author} {\bibfnamefont {T.}~\bibnamefont {Jacqmin}}, \bibinfo {author}
  {\bibfnamefont {A.}~\bibnamefont {Lema\^{\i}tre}}, \bibinfo {author}
  {\bibfnamefont {E.}~\bibnamefont {Galopin}}, \bibinfo {author} {\bibfnamefont
  {I.}~\bibnamefont {Sagnes}}, \bibinfo {author} {\bibfnamefont
  {A.}~\bibnamefont {Amo}}, \bibinfo {author} {\bibfnamefont {J.}~\bibnamefont
  {Bloch}}, \ and\ \bibinfo {author} {\bibfnamefont {E.}~\bibnamefont
  {Akkermans}},\ }\bibfield  {title} {\bibinfo {title} {\emph {Fractal Energy
  Spectrum of a Polariton Gas in a Fibonacci Quasiperiodic Potential}},\ }\href
  {\doibase 10.1103/PhysRevLett.112.146404} {\bibfield  {journal} {\bibinfo
  {journal} {Phys. Rev. Lett.}\ }\textbf {\bibinfo {volume} {112}},\ \bibinfo
  {pages} {146404} (\bibinfo {year} {2014})}\BibitemShut {NoStop}%
\bibitem [{\citenamefont {Zilberberg}(2021)}]{Zilberberg2021}%
  \BibitemOpen
  \bibfield  {author} {\bibinfo {author} {\bibfnamefont {O.}~\bibnamefont
  {Zilberberg}},\ }\bibfield  {title} {\bibinfo {title} {\emph {{Topology in
  quasicrystals}}},\ }\href {\doibase 10.1364/OME.416552} {\bibfield  {journal}
  {\bibinfo  {journal} {Opt. Mater. Express}\ }\textbf {\bibinfo {volume}
  {11}},\ \bibinfo {pages} {1143} (\bibinfo {year} {2021})}\BibitemShut
  {NoStop}%
\bibitem [{\citenamefont {Streda}(1982{\natexlab{a}})}]{Streda1982}%
  \BibitemOpen
  \bibfield  {author} {\bibinfo {author} {\bibfnamefont {P.}~\bibnamefont
  {Streda}},\ }\bibfield  {title} {\bibinfo {title} {\emph {Theory of quantised
  Hall conductivity in two dimensions}},\ }\href
  {http://stacks.iop.org/0022-3719/15/i=22/a=005} {\bibfield  {journal}
  {\bibinfo  {journal} {J. Phys. C: Solid State Physics}\ }\textbf {\bibinfo
  {volume} {15}},\ \bibinfo {pages} {L717} (\bibinfo {year}
  {1982}{\natexlab{a}})}\BibitemShut {NoStop}%
\bibitem [{\citenamefont {Streda}(1982{\natexlab{b}})}]{Streda1982bis}%
  \BibitemOpen
  \bibfield  {author} {\bibinfo {author} {\bibfnamefont {P.}~\bibnamefont
  {Streda}},\ }\bibfield  {title} {\bibinfo {title} {\emph {Quantised Hall
  effect in a two-dimensional periodic potential}},\ }\href
  {http://stacks.iop.org/0022-3719/15/i=36/a=006} {\bibfield  {journal}
  {\bibinfo  {journal} {J. Phys. C: Solid State Physics}\ }\textbf {\bibinfo
  {volume} {15}},\ \bibinfo {pages} {L1299} (\bibinfo {year}
  {1982}{\natexlab{b}})}\BibitemShut {NoStop}%
\bibitem [{\citenamefont {Dana}\ \emph {et~al.}(1985)\citenamefont {Dana},
  \citenamefont {Avron},\ and\ \citenamefont {Zak}}]{Dana1985}%
  \BibitemOpen
  \bibfield  {author} {\bibinfo {author} {\bibfnamefont {I.}~\bibnamefont
  {Dana}}, \bibinfo {author} {\bibfnamefont {Y.}~\bibnamefont {Avron}}, \ and\
  \bibinfo {author} {\bibfnamefont {J.}~\bibnamefont {Zak}},\ }\bibfield
  {title} {\bibinfo {title} {\emph {Quantised Hall conductance in a perfect
  crystal}},\ }\href {\doibase 10.1088/0022-3719/18/22/004} {\bibfield
  {journal} {\bibinfo  {journal} {J. Phys. C: Solid State Physics}\ }\textbf
  {\bibinfo {volume} {18}},\ \bibinfo {pages} {L679} (\bibinfo {year}
  {1985})}\BibitemShut {NoStop}%
\bibitem [{\citenamefont {Harper}(1955)}]{Harper1955}%
  \BibitemOpen
  \bibfield  {author} {\bibinfo {author} {\bibfnamefont {P.~G.}\ \bibnamefont
  {Harper}},\ }\bibfield  {title} {\bibinfo {title} {\emph {Single Band Motion
  of Conduction Electrons in a Uniform Magnetic Field}},\ }\href {\doibase
  10.1088/0370-1298/68/10/304} {\bibfield  {journal} {\bibinfo  {journal}
  {Proc. Phys. Soc. Sect. A}\ }\textbf {\bibinfo {volume} {68}},\ \bibinfo
  {pages} {874} (\bibinfo {year} {1955})}\BibitemShut {NoStop}%
\bibitem [{\citenamefont {Repellin}\ \emph {et~al.}(2020)\citenamefont
  {Repellin}, \citenamefont {L{\'{e}}onard},\ and\ \citenamefont
  {Goldman}}]{Repellin2020}%
  \BibitemOpen
  \bibfield  {author} {\bibinfo {author} {\bibfnamefont {C.}~\bibnamefont
  {Repellin}}, \bibinfo {author} {\bibfnamefont {J.}~\bibnamefont
  {L{\'{e}}onard}}, \ and\ \bibinfo {author} {\bibfnamefont {N.}~\bibnamefont
  {Goldman}},\ }\bibfield  {title} {\bibinfo {title} {\emph {{Fractional Chern
  insulators of few bosons in a box: Hall plateaus from center-of-mass drifts
  and density profiles}}},\ }\href {\doibase 10.1103/PhysRevA.102.063316}
  {\bibfield  {journal} {\bibinfo  {journal} {Phys. Rev. A}\ }\textbf {\bibinfo
  {volume} {102}},\ \bibinfo {pages} {063316} (\bibinfo {year}
  {2020})}\BibitemShut {NoStop}%
\bibitem [{\citenamefont {Janecek}\ \emph {et~al.}(2013)\citenamefont
  {Janecek}, \citenamefont {Aichinger},\ and\ \citenamefont
  {Hern{\'{a}}ndez}}]{Janecek2013}%
  \BibitemOpen
  \bibfield  {author} {\bibinfo {author} {\bibfnamefont {S.}~\bibnamefont
  {Janecek}}, \bibinfo {author} {\bibfnamefont {M.}~\bibnamefont {Aichinger}},
  \ and\ \bibinfo {author} {\bibfnamefont {E.~R.}\ \bibnamefont
  {Hern{\'{a}}ndez}},\ }\bibfield  {title} {\bibinfo {title} {\emph
  {{Two-dimensional Bloch electrons in perpendicular magnetic fields: An exact
  calculation of the Hofstadter butterfly spectrum}}},\ }\href {\doibase
  10.1103/PhysRevB.87.235429} {\bibfield  {journal} {\bibinfo  {journal} {Phys.
  Rev. B}\ }\textbf {\bibinfo {volume} {87}},\ \bibinfo {pages} {235429}
  (\bibinfo {year} {2013})}\BibitemShut {NoStop}%
\bibitem [{\citenamefont {Laughlin}(1981)}]{Laughlin1981}%
  \BibitemOpen
  \bibfield  {author} {\bibinfo {author} {\bibfnamefont {R.~B.}\ \bibnamefont
  {Laughlin}},\ }\bibfield  {title} {\bibinfo {title} {\emph {Quantized Hall
  conductivity in two dimensions}},\ }\href {\doibase 10.1103/PhysRevB.23.5632}
  {\bibfield  {journal} {\bibinfo  {journal} {Phys. Rev. B}\ }\textbf {\bibinfo
  {volume} {23}},\ \bibinfo {pages} {5632} (\bibinfo {year}
  {1981})}\BibitemShut {NoStop}%
\bibitem [{\citenamefont {Mochol-Grzelak}\ \emph {et~al.}(2018)\citenamefont
  {Mochol-Grzelak}, \citenamefont {Dauphin}, \citenamefont {Celi},\ and\
  \citenamefont {Lewenstein}}]{Mochol-Grzelak2018}%
  \BibitemOpen
  \bibfield  {author} {\bibinfo {author} {\bibfnamefont {M.}~\bibnamefont
  {Mochol-Grzelak}}, \bibinfo {author} {\bibfnamefont {A.}~\bibnamefont
  {Dauphin}}, \bibinfo {author} {\bibfnamefont {A.}~\bibnamefont {Celi}}, \
  and\ \bibinfo {author} {\bibfnamefont {M.}~\bibnamefont {Lewenstein}},\
  }\bibfield  {title} {\bibinfo {title} {\emph {{Efficient algorithm to compute
  the second Chern number in four dimensional systems}}},\ }\href {\doibase
  10.1088/2058-9565/aae93b} {\bibfield  {journal} {\bibinfo  {journal} {Quantum
  Sci. Technol.}\ }\textbf {\bibinfo {volume} {4}},\ \bibinfo {pages} {014009}
  (\bibinfo {year} {2018})}\BibitemShut {NoStop}%
\bibitem [{\citenamefont {de~Blas}\ and\ \citenamefont
  {Axel}(2004)}]{Blas2004}%
  \BibitemOpen
  \bibfield  {author} {\bibinfo {author} {\bibfnamefont {A.-Z.~E.}\
  \bibnamefont {de~Blas}}\ and\ \bibinfo {author} {\bibfnamefont
  {F.}~\bibnamefont {Axel}},\ }\bibfield  {title} {\bibinfo {title} {\emph
  {Diophantine equation for the 3D transport coefficients of Bloch electrons in
  a strong tilted magnetic field with quantum Hall effect}},\ }\href {\doibase
  10.1088/0953-8984/16/43/010} {\bibfield  {journal} {\bibinfo  {journal} {J.
  Phys.: Condens. Matter}\ }\textbf {\bibinfo {volume} {16}},\ \bibinfo {pages}
  {7673} (\bibinfo {year} {2004})}\BibitemShut {NoStop}%
\bibitem [{\citenamefont {Aidelsburger}(2015)}]{aidelsburger2015artifi}%
  \BibitemOpen
  \bibfield  {author} {\bibinfo {author} {\bibfnamefont {M.}~\bibnamefont
  {Aidelsburger}},\ }\href {\doibase 10.1007/978-3-319-25829-4} {\emph
  {\bibinfo {title} {Artificial gauge fields with ultracold atoms in optical
  lattices}}}\ (\bibinfo  {publisher} {Springer},\ \bibinfo {year}
  {2015})\BibitemShut {NoStop}%
\end{thebibliography}
\end{document}